\newif\ifconfver
\confvertrue        

\newif\ifonecoltab

\newif\ifplainver  

\ifplainver
    \confverfalse                         
\fi

\ifconfver
     \documentclass[10pt,journal]{IEEEtran}
\else
    \ifplainver
        \documentclass[11pt]{article}
        \usepackage{fullpage}
    \else
        \documentclass[11pt,draftcls,onecolumn]{IEEEtran}
    \fi
\fi

\usepackage{calc,amsfonts,amssymb,amsmath,bm,url,color,theorem,graphicx,cite,epstopdf,nicefrac,bbold,stmaryrd}
\usepackage{psfrag,subfigure,float}
\usepackage{tikztensor}
\usepackage{extramath}
\usepackage{tablefootnote}
\usepackage{bbding}

\usepackage[algoruled,linesnumbered]{algorithm2e}
\usepackage{multirow,framed}
\usepackage{mdframed}
\usepackage[normalem]{ulem}
\usepackage{xcolor} 
\usepackage{tikz}
\usepackage{threeparttable}
\usetikzlibrary{fit,calc,spy}
\usepackage{siunitx}

\colorlet{pink}{red!40}
\colorlet{bluelight}{cyan!60}

\definecolor{brightpink}{rgb}{1.0, 0.0, 0.5}
\newcommand{\nicolas}[1]{{\color{brightpink}#1}}

\usepackage{comment}

\usepackage{extramath}
\usepackage{tikztensor}
\usepackage{advancedpictures}

\usepackage{booktabs}
\usetikzlibrary{calc}

\tikzset{myarrow/.append style={->,shorten >=2pt,shorten <=2pt}}
\definecolor{kulblue}{RGB}{0,85,165}
\definecolor{mycolor1}{rgb}{0.00000,0.44700,0.74100}%
\tikzset{fancy/.style={2d,draw=kulblue!60!white,fill=kulblue!10!white}}

\newcommand{\Y}{\boldsymbol{Y}}
\newcommand{\G}{\boldsymbol{G}}

\newcommand{\X}{\boldsymbol{X}}

\renewcommand{\H}{\boldsymbol{H}}

\newcommand{\A}{\boldsymbol{A}}
\newcommand{\B}{\boldsymbol{B}}

\newcommand{\Z}{\boldsymbol{Z}}
\newcommand{\x}{\boldsymbol{x}}

\makeatletter
\let\@c@old=\c
\makeatother
\renewcommand{\c}{\boldsymbol{c}}

\renewcommand{\T}{{\!\top\!}}
\DeclareMathOperator*{\argmin}{\arg\min}

\definecolor{orange}{RGB}{255,107,0}

\def\diag{\mathrm{diag}}

\usepackage{xcolor}
\usepackage{psfrag,framed}
\usepackage{lipsum}
\usepackage{chapterbib}
\PassOptionsToPackage{normalem}{ulem}
\usepackage{ulem}
\definecolor{shadecolor}{RGB}{220,220,220}

\begin{document}

\newcommand{\papertitle}{
Computing Large-Scale Matrix and Tensor Decomposition with Structured Factors: A Unified Nonconvex Optimization Perspective
}

\newcommand{\paperabstract}{
The proposed magazine article aims at offering a comprehensive tutorial for the computational aspects of structured matrix and tensor factorization. Unlike existing tutorials that mainly focus on {\it algorithmic procedures} for a small set of problems, e.g., nonnegativity or sparsity-constrained factorization, our plan is to take a {\it top-down} approach: we will start with general optimization theory (e.g., inexact and accelerated block coordinate descent, stochastic optimization, and Gauss-Newton methods) that covers a wide range of factorization problems with diverse constraints and regularization terms of engineering interest.  Then, we will go `under the hood' to showcase specific algorithm design under these introduced principles. We will pay a particular attention to recent algorithmic developments in structured tensor and matrix factorization (e.g., random sketching and adaptive step size based stochastic optimization and structure-exploiting second-order algorithms), which are the state of the art---yet much less touched upon in the literature compared to {\it block coordinate descent} (BCD)-based methods. We expect that the article to have high educational values in the field of structured factorization and hope to stimulate more research in this important and exciting direction.
}


\ifplainver

    \date{\today}

    \title{\papertitle}

    \author{Xiao Fu, Nico Vervliet, Lieven De Lathauwer, Kejun Huang and Nicolas Gillis
    }

    \maketitle


\else
    \title{\papertitle}

    \ifconfver \else {\linespread{1.1} \rm \fi

\author{    Xiao Fu, Nico Vervliet, Lieven De Lathauwer, Kejun Huang and Nicolas Gillis

\thanks{X. Fu is supported by the National Science Foundation under projects ECCS-1608961, ECCS-1808159, and III-1910118, and the Army Research Office under projects ARO W911NF-19-1-0247 and ARO W911NF-19-1-0407. 
N. Vervliet is supported by a junior postdoctoral fellowship (12ZM220N) from the Research Foundation---Flanders (FWO). 
The work of the Belgian team is also supported by (1)~the Fonds de la Recherche Scientifique--FNRS and the Fonds Wetenschappelijk Onderzoek--Vlaanderen under EOS Project no 30468160 (SeLMA), (2)~KU Leuven Internal Funds C16/15/059 and ID-N project no 3E190402, and (3)~the Flemish Government (AI Research Program). 
N. Gillis acknowledges the support by the European Research Council (ERC starting grant no 679515).

X. Fu is with Oregon State University, Corvallis, OR 97331, USA; E-mail: xiao.fu@oregonstate.edu. 
N. Vervliet  and L. De Lathauwer are with KU Leuven, Leuven, Belgium; E-mail: (Nico.Vervliet, Lieven.DeLathauwer)@kuleuven.be.
K. Huang is with University of Florida, Gainesville, FL 32611, USA; E-Mail: kejun.huang@ufl.edu.
N. Gillis is with the University of Mons, Mons, Belgium; E-Mail: nicolas.gillis@umons.ac.be. 

}
}

    \maketitle

    \ifconfver \else
        \begin{center} \vspace*{-2\baselineskip}
        \end{center}
    \fi



    \ifconfver \else \IEEEpeerreviewmaketitle} \fi

 \fi

\ifconfver \else
    \ifplainver \else
        \newpage
\fi \fi

\section{Introduction}

In the past 20 years, low-rank tensor and matrix decomposition models (LRDMs) have become indispensable tools for signal processing, machine learning, and data science. 
LRDMs represent high-dimensional, multi-aspect, and multimodal data using low-dimensional latent factors in a succinct and parsimonious way.
LRDMs can serve for a variety of purposes, e.g., data embedding (dimensionality reduction), denoising, { 
latent variable analysis,}
model parameter estimation, and big data compression; see \cite{gillis2014and,sidiropoulos2017tensor,cichocki2015tensor,fu2018nonnegative,kolda2009tensor} for surveys of applications.

\begin{figure}[t]
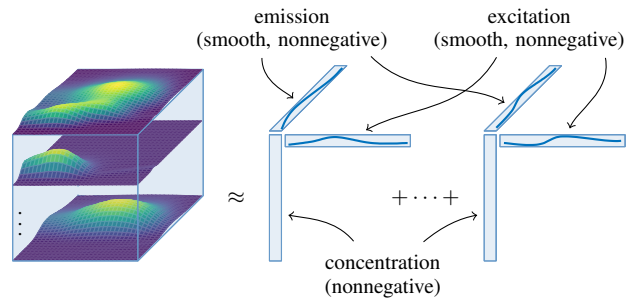

	\centering
        \setlength{\figurewidth}{4.35cm}
\setlength{\figureheight}{3.5cm}

\newsavebox\surfboxone
\begin{lrbox}{\surfboxone}
  \input{img/surf-amino1.tikz}
\end{lrbox}
\newsavebox\surfboxtwo
\begin{lrbox}{\surfboxtwo}
  \input{img/surf-amino3.tikz}
\end{lrbox}
\newsavebox\surfboxthree
\begin{lrbox}{\surfboxthree}
  \input{img/surf-amino4.tikz}
\end{lrbox}

\newsavebox\vectoronebox
\begin{lrbox}{\vectoronebox}
  \begin{tikzpicture}
    \begin{axis}[%
      width=11.5mm,
      height=0.8mm,
      rotate=45,
      scale only axis,
      xmin=0,
      xmax=1,
      ymin=0,
      ymax=1,
      axis y line=none,
      axis x line=none,
      clip=false,
      ]
      \addplot [color=mycolor1,solid,line width=0.8pt,forget plot]
      table[row sep=crcr]{%
0.00000 0.01490\\
0.00651 0.05912\\
0.02555 0.16372\\
0.04113 0.25940\\
0.05672 0.35125\\
0.07342 0.45240\\
0.09307 0.54737\\
0.11382 0.63633\\
0.13456 0.70996\\
0.15419 0.77266\\
0.17430 0.83203\\
0.19676 0.89257\\
0.21747 0.92504\\
0.23818 0.94601\\
0.25888 0.95933\\
0.27957 0.96690\\
0.30026 0.95532\\
0.32094 0.94374\\
0.34161 0.92259\\
0.36228 0.89953\\
0.38295 0.87071\\
0.40362 0.83999\\
0.42429 0.80735\\
0.44495 0.77279\\
0.46561 0.73440\\
0.48628 0.69602\\
0.50694 0.65572\\
0.52760 0.61733\\
0.54826 0.57512\\
0.56892 0.53290\\
0.58957 0.48686\\
0.61024 0.44847\\
0.63089 0.40434\\
0.65155 0.36021\\
0.67221 0.31799\\
0.69287 0.27769\\
0.71353 0.23548\\
0.73419 0.19326\\
0.75486 0.15871\\
0.77552 0.12032\\
0.79619 0.09342\\
0.81686 0.06844\\
0.83754 0.04921\\
0.85821 0.03380\\
0.87890 0.02988\\
0.89959 0.02979\\
0.92029 0.03928\\
0.94099 0.06408\\
0.96170 0.08123\\
0.98238 0.07540\\
0.99790 0.07533\\
};
    \end{axis}
  \end{tikzpicture}
\end{lrbox}

\newsavebox\vectortwobox
\begin{lrbox}{\vectortwobox}
  \begin{tikzpicture}
    \begin{axis}[%
      width=16mm,
      height=1.4mm,
      scale only axis,
      xmin=0,
      xmax=1,
      ymin=0,
      ymax=1,
      axis y line=none,
      axis x line=none
      ]
      \addplot [color=mycolor1,solid,line width=0.8pt,forget plot]
      table[row sep=crcr]{%
0.00655 0.07494\\
0.02236 0.08551\\
0.03817 0.11085\\
0.05397 0.12635\\
0.06978 0.15169\\
0.08559 0.17211\\
0.10139 0.19745\\
0.11720 0.21787\\
0.13300 0.24321\\
0.14881 0.27347\\
0.16462 0.29389\\
0.18042 0.32415\\
0.19623 0.34949\\
0.21204 0.38468\\
0.22784 0.42478\\
0.24365 0.47473\\
0.25946 0.51484\\
0.27526 0.56971\\
0.29107 0.62458\\
0.30688 0.66469\\
0.32268 0.70972\\
0.33849 0.74490\\
0.35429 0.77024\\
0.37010 0.78574\\
0.38591 0.79631\\
0.40171 0.79704\\
0.41752 0.78301\\
0.43333 0.76405\\
0.44913 0.74509\\
0.46494 0.71629\\
0.48075 0.68257\\
0.49655 0.64885\\
0.51236 0.61512\\
0.52816 0.57648\\
0.54397 0.54275\\
0.55978 0.49919\\
0.57558 0.46546\\
0.59139 0.42682\\
0.60720 0.38817\\
0.62300 0.35445\\
0.63881 0.32073\\
0.65462 0.29193\\
0.67042 0.26805\\
0.68623 0.24417\\
0.70203 0.22521\\
0.71784 0.21117\\
0.73365 0.20206\\
0.74945 0.19295\\
0.76526 0.17891\\
0.78107 0.17964\\
0.79687 0.18037\\
0.81268 0.18110\\
0.82849 0.18183\\
0.84429 0.18256\\
0.86010 0.18330\\
0.87590 0.18403\\
0.89171 0.18476\\
0.90752 0.18549\\
0.92332 0.19606\\
0.93913 0.20664\\
0.95494 0.21721\\
0.97074 0.21794\\
0.98655 0.22852\\
0.99709 0.21916\\
};
    \end{axis}
  \end{tikzpicture}
\end{lrbox}

\newsavebox\vectorthreebox
\begin{lrbox}{\vectorthreebox}
  \begin{tikzpicture}
    \begin{axis}[%
      width=11.5mm,
      height=0.8mm,
      rotate around={45:(current axis.origin)},
      scale only axis,
      xmin=0,
      xmax=1,
      ymin=0,
      ymax=1,
      axis y line=none,
      axis x line=none,
      clip=false
      ]
      \addplot [color=mycolor1,solid,line width=0.8pt,forget plot]
      table[row sep=crcr]{%
0.00736 0.03896\\
0.02253 0.03941\\
0.03769 0.03986\\
0.05285 0.04031\\
0.06801 0.04077\\
0.08317 0.04122\\
0.09834 0.04167\\
0.11350 0.04212\\
0.12870 0.05269\\
0.14388 0.05820\\
0.15906 0.06371\\
0.17425 0.07427\\
0.18945 0.08484\\
0.20467 0.10046\\
0.21988 0.11609\\
0.23512 0.13677\\
0.25036 0.15998\\
0.26686 0.18133\\
0.28086 0.21020\\
0.29619 0.25996\\
0.31157 0.32110\\
0.32699 0.39489\\
0.34242 0.46994\\
0.35786 0.55005\\
0.37328 0.62257\\
0.38872 0.70015\\
0.40411 0.76508\\
0.41946 0.81864\\
0.43479 0.86714\\
0.45015 0.92449\\
0.46541 0.95023\\
0.48061 0.96332\\
0.49579 0.96883\\
0.51094 0.96423\\
0.52605 0.95204\\
0.54115 0.93479\\
0.55622 0.90742\\
0.57128 0.88006\\
0.58634 0.85269\\
0.60138 0.81774\\
0.61737 0.76562\\
0.63015 0.73832\\
0.64742 0.69066\\
0.66043 0.65464\\
0.67141 0.61628\\
0.68774 0.58832\\
0.70049 0.54963\\
0.71245 0.50296\\
0.72549 0.47452\\
0.73745 0.42785\\
0.75049 0.39941\\
0.76245 0.35274\\
0.77657 0.34254\\
0.79055 0.29441\\
0.80251 0.24773\\
0.81532 0.22801\\
0.83162 0.19057\\
0.84669 0.16510\\
0.86175 0.13774\\
0.87681 0.11037\\
0.89190 0.09059\\
0.90700 0.07334\\
0.92212 0.06115\\
0.93725 0.05149\\
0.95241 0.05194\\
0.96759 0.05745\\
0.98280 0.07307\\
};
    \end{axis}
  \end{tikzpicture}
\end{lrbox}

\newsavebox\vectorfourbox
\begin{lrbox}{\vectorfourbox}
  \begin{tikzpicture}
    \begin{axis}[%
      width=16mm,
      height=1.4mm,
      scale only axis,
      xmin=0,
      xmax=1,
      ymin=0,
      ymax=1,
      axis y line=none,
      axis x line=none,
      clip=false,
      ]
      \addplot [color=mycolor1,solid,line width=0.8pt,forget plot]
      table[row sep=crcr]{%
0.01129 0.30253\\
0.02813 0.29168\\
0.04501 0.26908\\
0.06189 0.24649\\
0.07877 0.22389\\
0.09565 0.20129\\
0.11252 0.18261\\
0.12938 0.16393\\
0.14624 0.14917\\
0.16309 0.13441\\
0.17993 0.12356\\
0.19677 0.11272\\
0.21356 0.11362\\
0.23036 0.11453\\
0.24716 0.11543\\
0.26392 0.12417\\
0.28068 0.13683\\
0.29738 0.16515\\
0.31403 0.20523\\
0.33205 0.25811\\
0.34724 0.31475\\
0.36369 0.41358\\
0.38015 0.50849\\
0.39663 0.59949\\
0.41309 0.69440\\
0.42822 0.76867\\
0.44614 0.84701\\
0.46281 0.88317\\
0.47948 0.91933\\
0.49625 0.92807\\
0.51305 0.92897\\
0.52985 0.92988\\
0.54666 0.92687\\
0.56349 0.91602\\
0.58035 0.90126\\
0.59720 0.88650\\
0.61405 0.87173\\
0.63093 0.84914\\
0.64781 0.82654\\
0.66469 0.80394\\
0.68157 0.78135\\
0.69845 0.75875\\
0.71533 0.73616\\
0.73223 0.70964\\
0.74911 0.68705\\
0.76600 0.66053\\
0.78288 0.63794\\
0.79976 0.61534\\
0.81664 0.59274\\
0.83352 0.57015\\
0.85040 0.54755\\
0.86726 0.53279\\
0.88411 0.51802\\
0.90095 0.50718\\
0.91779 0.49633\\
0.93461 0.48940\\
0.95142 0.48639\\
0.96819 0.49513\\
0.98493 0.51170\\
};
    \end{axis}
  \end{tikzpicture}
\end{lrbox}

\begin{tikzpicture}[img/.style={inner sep=0pt}, vec/.style={mode-3 vector, 2d,
    width=0.25, dim={2},anchor=A,yshift=0.3mm,text fraction=0.95,text
    offset=-0.29mm,
    draw=kulblue!60!white,fill=kulblue!10!white},
    node distance=2mm,
    fancy/.style={2d,draw=kulblue!60!white,fill=kulblue!10!white},
    font=\footnotesize,tensor scale=0.8,2d]

  \node [tensor,fancy,dim={2.1,2.08,2.82},only back faces,dashed back lines] (T) {};
  \node at (T.B) [anchor=south west,inner sep=0pt] {\usebox\surfboxone};
  \node at (T.B) [anchor=south west,inner sep=0pt,yshift=1cm] {\usebox\surfboxtwo};

  \node [tensor,fancy,dim={2.1,2.08,2.82},fill=kulblue!60!white,fill opacity=0.2] (T) {};
  \node at (T.A) [anchor=south west,inner sep=0pt] {\usebox\surfboxthree};

  \node [right=of T] (eq) {$\approx$};
  \node [right=of eq,tensor,dim={2.1,0.2,0},fancy] (a1) {};  
  \node at (a1.A) [mode-3 vector,dim={2.82},fancy,anchor=A,yshift=0.5mm,text
  fraction=0.98,text offset=-1.35] (c1)
  {\usebox\vectoronebox};
  \node at (a1.C) [tensor,dim={0.2,2.08,0},fancy,anchor=A,xshift=0.5mm] (b1)
  {\usebox\vectortwobox};

  \coordinate [right=of b1] (tmp);
  \node at (tmp|-eq) (eq) {$+\cdots+$};
  \node [right=of eq,tensor,dim={2.1,0.2,0},fancy] (a2) {};  
  \node at (a2.A) [mode-3 vector,dim={2.82},fancy,anchor=A,yshift=0.5mm,text
  fraction=0.98,text offset=-1.35] (c2)
  {\usebox\vectorthreebox};
  \node at (a2.C) [tensor,dim={0.2,2.08,0},fancy,anchor=A,xshift=0.5mm] (b2)
  {\usebox\vectorfourbox};

  \node at (a1.A) [text width=4cm,align=center,yshift=14mm,xshift=3mm] (tc)
  {emission\\(smooth, nonnegative)};

  \node at (a2.A) [text width=2.6cm,align=center,yshift=14mm,xshift=6mm] (tb)
  {excitation\\(smooth, nonnegative)};

  \node at ($(a1.D)!0.5!(a2.B)$) [text width=4cm,align=center,yshift=-2mm] (ta)
  {concentration\\(nonnegative)};
    
  \draw [myarrow,->,shorten >=5pt] (ta) to [out=130,in=-10] ($(a1)+(0,-2mm)$);
  \draw [myarrow,->,shorten >=5pt] (ta) to [out=40,in=190] ($(a2)+(0,-2mm)$);

  \draw [myarrow,->,shorten >=8pt] ($(tc.south)+(-4mm,1mm)$) to [out=-90,in=135] ($(c1)+(0,-2mm)$);
  \draw [myarrow,->,shorten >=8pt] ($(tc.south)+(10mm,1mm)$) to [out=-45,in=150] ($(c2)+(0,-2mm)$);

  \draw [myarrow,->,shorten >=8pt] ($(tb.south)+(-4mm,1mm)$) to [out=-120,in=30] ($(b1)+(0,0mm)$);
  \draw [myarrow,->,shorten >=8pt] ($(tb.south)+(10mm,1mm)$) to [out=-80,in=70] ($(b2)+(0,-1mm)$);

  \node at (T.B) [xshift=1mm,yshift=6mm] {$\vdots$};
  
\end{tikzpicture}
		\caption{{ An SLRD model} in fluorescence data analytics. The rank-one components correspond to different analytes constituting the data samples. The latent factors have a physical meaning, and 
		using prior structural information improves decomposition performance.}
		\label{fig:flo}

\end{figure}

\begin{figure}[t]
	\centering
           \input{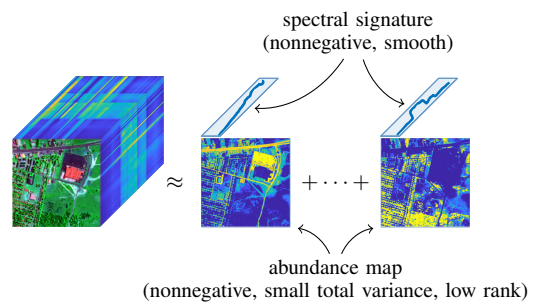}
		\caption{ The linear mixture model for hyperspectral unmixing (HU). The HU problem can be either considered as a nonnegative matrix factorization problem \cite{gillis2014and,fu2018nonnegative} or a block-term tensor decomposition problem \cite{qian2017matrix}. Both are SLRDs.}
		\label{fig:hu}
\end{figure}

\begin{figure}[t]
	\centering
		\input{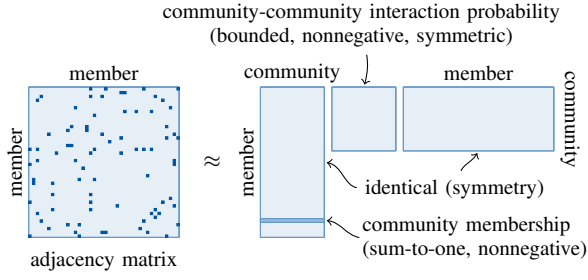}
		\caption{ An SLRD perspective for community detection under the mixed membership stochastic blockmodel \cite{huang2019detecting}. 
		The binary adjacency matrix can be considered as a noisy structured low-rank model.
		Again, a series of model priors can be used as structural constraints on the latent factor matrices on the right hand side.}
		\label{fig:cd}

\end{figure}

LRDM often poses challenging optimization problems.
This article aims at introducing the recent advances and key computational aspects in {\it structured low-rank matrix and tensor decomposition} (SLRD). Here, ``structured decomposition'' refers to the techniques that impose structural requirements (e.g., nonnegativity, smoothness, and sparsity) onto the latent factors when computing the decomposition (see { Figs.~\ref{fig:flo}-\ref{fig:cd} for a number of examples and the references therein}). Incorporating structural information is well-motivated in many cases. For example, adding constraints/regularization terms typically enhances performance in the presence of noise and modeling errors, since constraints and regularization terms impose prior information on the latent factors. For certain tensor decompositions like the {\it canonical polyadic decomposition} (CPD), adding constraints (such as nonnegativity or orthogonality) converts ill-posed optimization problems (where optimal solutions do not exist) into well-posed ones \cite{lim2009nonnegative}. In addition, constraints and regularization terms can make the results more ``interpretable''; e.g., if one aims at estimating probability mass functions (PMFs) or power spectra from data, adding probability simplex or nonnegativity constraints to the latent factors makes the outputs consistent with the design objectives. 
For matrix decomposition, adding constraints is even more critical---e.g., adding nonnegativity to the latent factors can make highly nonunique matrix decompositions have essentially unique latent factors \cite{gillis2014and,fu2018nonnegative}---as model uniqueness is a core consideration in parameter identification, signal separation, and unsupervised machine learning.


Due to the importance of LRDMs, a plethora of algorithms have been proposed. 
The overview papers on tensor decomposition \cite{kolda2009tensor,sidiropoulos2017tensor} have 
{  
discussed}
many relevant models, their algebraic properties, and popular decomposition algorithms (without emphasizing on structured decomposition).
\color{black} 
In terms of incorporating structural information, nonnegativity and sparsity-related algorithms have been given the most attention, due to their relevance in image, video and text data analytics; see, e.g., the tutorial articles published in 2014 \cite{zhou2014nonnegative} and \cite{gillis2014and} for LRDMs with nonnegativity constraints.

In this article, instead of offering a comprehensive overview of algorithms under different low-rank decomposition models or particular types of constraints, 
we provide a unified and principled nonconvex nonsmooth optimization perspective for SLRD. We will pay particular attention to the following two aspects. First, we will introduce how different nonconvex optimization tools (in particular, block coordinate descent, Gauss--Newton algorithms, and stochastic optimization) can be combined with tensor/matrix structures to come up with lightweight algorithms while considering various structural requirements. 
Second, we will touch upon the key considerations for ensuring that these algorithms have convergence guarantees (e.g., guarantees for convergence to a stationary point), since convergence guarantees are important for designing stable and disciplined algorithms.
Both nonconvex nonsmooth optimization and tensor/matrix decomposition are nontrivial. We hope that this article could entail the readers (especially graduate students) an entry point for understanding the key ingredients that are needed for designing structured decomposition algorithms---in a disciplined way.

\smallskip

\noindent
{\bf Notation.} We follow the established conventions in signal processing, and use ${\cal T}$, $\X$ and $\x$ to denote a tensor, a matrix and a vector, respectively. The notations $\otimes$, $\odot$, $\circledast$, and $\circ$ denote the Kronecker product, Khatri--Rao product, Hadamard product, and outer product, respectively. The matlab notation $\X(m,:)$ is used to denote the $m$th row of $\X$, and other matlab notations such as $\X(:,n)$ and $\X(i,j)$ are also used. 
In some cases, $[\X]_{i,j}$ and $[\x]_j$ denote the $(i,j)$th entry of $\X$ and the $j$th element of $\x$, respectively. 
The notation $\X=[\X_1;\ldots;\X_N]=[\X_1^\T,\ldots,\X_2^\T]^\T$ denotes the concatenation of the matrices $\{\X_i\}_{n=1}^N$. 

\section{Problem Statement}

\subsection{Low-rank Matrix and Tensor Decomposition Models}

Under a noiseless setting, matrix decomposition aims at finding the following representation of a data matrix $\X$:
\begin{equation}\label{eq:mf}
 \X =\A_1\A_2^\T=\sum_{r=1}^R\A_1(:,r)\circ \A_2(:,r), 
\end{equation}       
where $\X\in\mathbb{R}^{I_1\times I_2}$, $\A_1\in\mathbb{R}^{I_1\times R}$ and $\A_2\in\mathbb{R}^{I_2\times R}$. The integer $R\leq \min \{I_1,I_2  \}$ is the smallest integer such that the equality above holds---$R$ denotes the matrix rank.
If the data entries have more than two indices, the data array is called a {\it tensor}. 
Unlike matrices whose rank decomposition is defined as in \eqref{eq:mf}, there are a variety of tensor decomposition models involving different high-order generalizations of matrix rank. One of the most popular models is CPD \cite{harshman1970foundations}.
For an $N$th-order tensor ${\cal T}\in\mathbb{R}^{I_1 \times \ldots\times I_N}$, its CPD representation is as follows:
\begin{equation}\label{eq:cpd}
{\cal T}=\sum_{r=1}^R \A_1(:,r)\circ\ldots \circ\A_N(:,r) \eqqcolon \llbracket\A_1,\ldots,\A_N\rrbracket,
\end{equation}
where $R$ is again the smallest integer such that the equality holds (i.e., $R$ is the CP rank of ${\cal T}$), and $\A_n\in\mathbb{R}^{I_n\times R}$ denotes the mode-$n$ latent factor (see the visualization of a third-order case in Fig.~\ref{fig:flo}).
Besides CPD, {there is, for instance, the Tucker decomposition model}, i.e.,
${\cal T}={\cal G}\times_1 \A_1 \times_2 \ldots \times _N \A_N,$
where ${\cal G}$ denotes the so-called core tensor, and $\times_n$ is the mode-$n$ product.
More recently, a series of extensions and hybrid models have also emerged, including the \textit{block-term decomposition} (BTD), \textit{multilinear rank-$(L_r,L_r,1)$ decomposition} ({LL1}), { coupled CPD/BTD models (see the insert ``{\bf Handling Special Constraints via Parameterization}'' and references therein), the tensor train model, and the hierarchical Tucker model; see \cite{sidiropoulos2017tensor} and the references therein.}
In this article, we will mainly focus on the models in \eqref{eq:mf} and \eqref{eq:cpd}, and use them to illustrate different algorithm design principles.
Generalization to other models will also be briefly discussed at the end.

{ In their general formulation,}
most SLRD problems are NP-hard \cite{hillar2013most,gillis2018complexity,vavasis2009complexity}.
Apart from that, the era of big data brings its own challenges. 
\color{black}
For example, a $2000\times 2000\times 2000$ tensor (i.e., $I_n=I=2000$ for all $n$) requires 58\,GB memory (if the double precision is used). Already when there is no constraint or regularization on the latent factors, using first-order optimization techniques (e.g., gradient descent or block coordinate descent) under the ``optimization-friendly" Euclidean loss costs $\order{RI^3}$ floating point operations (flop) per iteration for the rank-$R$ CPD of this tensor. With constraints and regularization terms, the complexity might be higher. The situation gets worse when one deals with higher-order tensors.
Hence designing effective algorithms requires synergies between sophisticated optimization tools and the algebraic structures embedded in LRDMs.

\subsection{Structured Decomposition as Nonconvex Optimization}

SLRD can be viewed from a model fitting perspective. That is, we hope to find a tensor/matrix model that best approximates the data tensor or matrix under a certain distance measure, with prior information about the model parameters. 
This point of view makes a lot of sense. In practice, the data matrix/tensor often consists of low-rank ``essential information'' and high-rank noise---and thus using a model fitting formulation instead of seeking an exact decomposition as in \eqref{eq:mf} or \eqref{eq:cpd} is more meaningful.
Conceptually, the SLRD problems can be summarized as follows: 
\begin{mdframed}[backgroundcolor=blue!10,nobreak=true] \vspace{-0.2cm}
\begin{align}\label{eq:problem}
&\min_{\text{model~param.}}~{\rm dist}\left(\text{data},  \text{model}\right) + \left(\hspace{-0.2cm}{\footnotesize \begin{array}{c}
\text{penalty~for}\\\text{structure~violation}
\end{array} }\hspace{-0.2cm}\right) \nonumber \\
&\text{under}~~~\text{structural~constraints,}
\end{align}
\end{mdframed}  
where ${\rm dist}\left(  \X, \Y \right)$ is a ``distance measure'' { between $\X$ and $\Y$ in a certain sense}.
{ The most commonly used measure is the (squared) Euclidean distance, i.e., $${\rm dist}\left(  \X, \Y   \right)=\|\X-\Y\|_{\rm F}^2.$$
In addition, a number of other measures are of interest in data science. For example, the {\it Kullback--Leibler} (KL) divergence is often used 
for measuring the ``distance'' between distributions of random variables, and it is also commonly used in integer data fitting problems (since it is closely related to the maximum likelihood estimators (MLEs) that are associated with discrete RVs, e.g., those following the Poisson or Bernoulli distributions). The $\ell_1$ norm, the Huber function, and their nonconvex counterparts (e.g., the $\ell_p$ function where $0<p<1$ \cite{fu2016robust}) are used for outlier-robust data analytics; see more discussions in Section~\ref{sec:morediscuss}.} The ``structural constraints'' and ``structure violation penalty'' are imposed upon the model parameters [e.g., the $\A_n$'s in \eqref{eq:cpd}].
For example, consider CPD under sparsity and nonnegativity considerations, which finds applications in many data analytics problems \cite{beutel2014flexifact}:
\begin{align}\label{eq:constrained_cpd}
	\min_{\{\A_n\}_{n=1}^N}&~ \frac{1}{2}\left\|{\cal T} - \llbracket\A_1,\ldots,\A_N\rrbracket\right\|_{\rm F}^2  +\lambda \sum_{n=1}^N\|\A_n\|_1 \nonumber\\
	 {\rm s.t.}&~\A_n\geq \bm 0. 
\end{align}
From an optimization viewpoint, these SLRD problems can be summarized in a succinct form:
\begin{mdframed}[backgroundcolor=blue!10]
\begin{equation}\label{eq:constrained}
\min_{\bm \theta}~{f(\bm \theta)} + {h(\bm \theta)},
\end{equation}
\end{mdframed}
where $\bm \theta$ collects all the latent parameters of the tensor/matrix model of interest, $f(\bm \theta)$ represents the data fitting part, 
and $h(\bm \theta)$ represents regularization terms added on the latent factor. 
Note that the expression in \eqref{eq:constrained} also includes the case where $\bm \theta$ is subject to hard constraints; i.e., $\bm \theta\in{\cal C}$ can also be expressed as a penalty term, where $h(\bm \theta)$ is the indicator function of the set ${\cal C}$.
For example, in Problem~\eqref{eq:constrained_cpd}, $\bm \theta = [\bm \theta_1; \ldots; \bm \theta_N]$ where $\bm \theta_n = {\rm vec}(\A_n)$. Here, $h(\bm \theta)=\sum_{n=1}^N h_n(\bm \theta_n)$, and $h_n(\bm \theta_n)=h_n^{(1)}(\bm \theta_n)+h_n^{(2)}(\bm \theta_n)$---in which $h_n^{(1)}$ is the indicator function of the nonnegativity orthant
and $h_n^{(2)}(\bm \theta_n)$ the $L_1$ regularization.

Several observations can be made on the model~\eqref{eq:constrained_cpd}.
First, the SLRD problems are usually nonconvex, since the model approximation part $f(\bm \theta)$ is nonconvex;
in some cases, $h(\bm \theta)$ is also nonconvex; see, e.g., volume minimization-based NMF \cite{fu2018nonnegative}.
Second, the objective function in \eqref{eq:constrained} is oftentimes nonsmooth, especially when a non-differentiable regularization term $h(\bm \theta)$ is involved (e.g., an indicator function for enforcing ``hard constraints'' or an $L_1$ norm regularization term).
Like many nonconvex nonsmooth optimization problems, the SLRD problems are NP-hard in most cases~\cite{hillar2013most,vavasis2009complexity}. For general nonconvex optimization algorithms, the analytical tool for characterizing their global optimality-attaining properties has been elusive.
The convention from the optimization literature is to characterize the algorithms' stationary point-approaching properties, since $\bm \theta$ being a stationary point of \eqref{eq:constrained} is a necessary condition for $\bm \theta$ being an optimal solution.
Simply speaking, assume that the data fitting part $f(\bm \theta)$ is differentiable and ${\rm dom} (f+h)=\mathbb{R}^{d}$ where $d$ is the number of variables. Denote $F(\bm \theta)=f(\bm \theta)+h(\bm \theta)$. Then, any stationary point of Problem~\eqref{eq:constrained} satisfies the following:
\begin{equation}\label{eq:convergence}
 \bm 0 \in \partial F(\bm \theta) = \nabla f(\bm \theta ) +\partial h(\bm \theta),  
\end{equation}
where $\partial h(\bm \theta)$ denotes the limiting Fr\'echet subdifferential of $h(\cdot)$, which is the subgradient when $h(\cdot)$ is convex~\cite{xu2013block,xu2015block,razaviyayn2013unified}.

\section{BCD-based Approaches}

One of the workhorses for LRDMs is \textit{block coordinate descent} (BCD). 
The rationale behind BCD-based structured factorization is straightforward: The factorization problems with respect to (w.r.t.) a single block variable $\A_n$ in \eqref{eq:constrained_cpd} is convex under various models and $F(\cdot)$'s. 
BCD alternatingly updates the parameters $\bm \theta_n$ (the $n$th block of $\bm \theta$) while fixing the others: $\bm \theta_n^{(t+1)}$ is updated using  
\begin{equation}\label{eq:bcd}
 \argmin_{\bm \theta_n} f\left(\bm \theta_1^{(t+1)}, \ldots,\bm\theta_{n-1}^{(t+1)},\bm \theta_n,\bm\theta_{n-1}^{(t)},\ldots \bm \theta_N^{(t)} \right)  + h_n(\bm \theta_n),
\end{equation}
where $h_n(\cdot)$ is the part of $h(\cdot)$ that is imposed onto $\bm \theta_n$, and $\bm \theta^{(t)}$ denotes the optimization variables in iteration $t$.
In the sequel, we will use the shorthand notation
$  f(\bm \theta_n; \bm \theta^{(t)}_{-n}) =        f(\bm \theta_1^{(t+1)}, \ldots,\bm\theta_{n-1}^{(t+1)},\bm \theta_n,\bm\theta_{n-1}^{(t)},\ldots \bm \theta_N^{(t)} )$.

BCD and LRDMs are linked together through the ``matrix unfolding'' operation. Unfolding is a way of rearranging the elements of a tensor to a matrix.
The mode-$n$ unfolding (matricization) of $\cal T$ is as follows\cite{sidiropoulos2017tensor}\footnote{Note that tensor unfolding admits several forms in the literature. For example, the unfolding expressions in the two tutorial papers \cite{sidiropoulos2017tensor} and \cite{kolda2009tensor} are different. In this article, we follow the convention in \cite{sidiropoulos2017tensor}.}: 
for all $i_1,\ldots,i_N$, 
$$
\X_n(j,i_n)={\cal T}(i_1,\ldots,i_N),
$$
where $j=1+\sum_{\ell=1,\ell\neq n}(i_\ell-1)J_\ell$, $J_\ell=\prod_{m=1,m\neq n}^{\ell-1}I_m$.
For tensors with CP rank $R$, the unfolding has the following compact and elegant expression:
\begin{align}
  \label{eq:matrix-unfolding}
  \X_{n} =\H_{n}\A_n^\T,
\end{align}
where the matrix $\H_{n}\in\mathbb{R}^{(\prod_{\ell=1,\ell\neq n}^N I_n)\times R}$ is defined as:
\begin{align*}
\H_n &= \A_N \odot \ldots \odot \A_{n+1} \odot \A_{n-1} \odot \ldots \odot \A_1. 
\end{align*}
The readers are referred to \cite{sidiropoulos2017tensor,kolda2009tensor} for details of unfolding.
The unfolding operation explicitly ``pushes'' the latent factors to the rightmost position in the unfolded tensor representation---which helps efficient algorithm design.
Note that many tensor factorization models, e.g., Tucker, BTD, and LL1, have similar multilinearity properties in their respective unfolded representations \cite{de2008decompositions_3}.  
Representing the tensor using matrix unfolding, the BCD algorithm for structured CPD consists of the following updates in a cyclical manner:
\begin{align}\label{eq:cls}
	\A_n &\leftarrow\arg\min_{\A_n}~\frac{1}{2}\|\X_{n} - \H_n^{(t)}\A^\T  \|_{\rm F}^2 + h_n(\A_n),
\end{align}
where $\H_n^{(t)}= \A_N^{(t)} \odot \ldots \odot \A_{n+1}^{(t)} \odot \A_{n-1}^{(t+1)} \odot \ldots \odot \A_1^{(t+1)} $, since $\A_\ell$ for $\ell<n$ has been updated.

\subsection{Classic BCD based Structured Decomposition}
If $h_n(\A_n)$ is absent, Problem \eqref{eq:cls} admits an analytical solution, i.e., $\A_n^{(t+1)}\leftarrow (\H_n^{(t)})^\dag \X_n,$ which recovers the classic \emph{alternating least squares} (ALS) algorithm for the unconstrained least squares loss based CPD \cite{harshman1970foundations}.
In principle, if $h_n(\A_n)$ is convex, then any off-the-shelf convex optimization algorithms can be utilized to solve~\eqref{eq:cls}. However, in the context of SLRD, the algorithms employed should strike a good balance between complexity and efficiency. The reason is that~\eqref{eq:cls} can have a very large size, because  the row size of $\H_n^{(t)}$ is $\prod_{\ell=1,\ell\neq n}^N I_\ell$---which can reach millions even when $I_n$ is small. 

First-order optimization algorithms (i.e., optimization algorithms only using the gradient information) are known to be scalable, and thus are good candidates for handling~\eqref{eq:cls}. 
Proximal/projected gradient descent (PGD)~\cite{parikh2013proximal,lin2007projected} is perhaps the easiest to implement. PGD solves Problem~\eqref{eq:cls} using the following iterations:
$$\A_n^{(k+1)}\leftarrow  {\sf Prox}_{h_n}\left( \A_n^{(k)}  - \alpha \nabla_{\A_n} f\left(\A_n^{(k)};\A_{-n}^{(t)} \right) \right),$$
where $k$ indexes the iterations of the PGD algorithm. 
The notation ${\sf Prox}_h(\bm Z)$ is defined as
\[      {\sf Prox}_{h}\left( \bm Z \right) = \arg\min_{\bm Y}~h(\bm Y)+\frac{1}{2}\|\bm Y-\bm Z\|_{\rm F}^2,   \]
and  $\nabla_{\A_n} f(\A^{(k)},\A_{-n}^{(t)}) = \A^{(k)}(\H_n^{(t)})^\T\H_n^{(t)}  - \X_{n}^\T\H_n^{(t)}$. 
For a variety of $h(\cdot)$'s, the proximal operator is easy to compute. For example, if $h(\bm Z)=\lambda\|\Z\|_1$, we have
$ [{\sf Prox}_{\lambda \|\cdot\|_1}\left( \bm Z \right)]_{ij} = {\rm sign}(Z_{ij})(|Z_{ij}-\lambda|)_+,$
and if $h(\bm Z)$ is the indicator function of a closed set $\mathbb{H}$, then the proximal operator becomes a projection operator, i.e., 
${\sf Prox}_{h}\left( \bm Z \right) = {\sf Proj}_{\mathbb{H}}(\bm Z)=\arg\min_{\bm Y\in\mathbb{H}}~\|\bm Y-\bm Z\|_{\rm F}^2$. A number of $h(\cdot)$'s that admit simple proximal operations (e.g., $L_1$-norm) can be found in \cite{parikh2013proximal}. 

PGD is easy to implement when the proximal operator is simple. When the regularization is complicated, then using algorithms such as the alternating directional method of multipliers (ADMM) to replace PGD may be more effective; see \cite{huang2016flexible} for a collection of examples of ADMM-based constrained least squares solving.
Beyond PGD and ADMM, many other algorithms have been employed for handling the subproblem in \eqref{eq:cls} for different structured decomposition problems. For example, accelerated PGD, active set, and mirror descent have all been considered in the literature; see, e.g., \cite{guan2012nenmf,kim2008nonnegative}.

\subsection{Inexact BCD} \label{sec:inexactBCD}

Using off-the-shelf solvers to handle the subproblems under the framework of BCD is natural for many LRDMs.  
However, when the block variables $\bm \theta_{-n}$ is only roughly estimated, it is not necessarily efficient to exactly solve the subproblem 
w.r.t.\ $\bm \theta_n$---after all, $\bm \theta_{-n}$ will change in the next iteration.
This argument leads to a class of algorithms that solve the block subproblems in an inexact manner \cite{xu2013block,razaviyayn2013unified}.

Instead of directly minimizing $f(\A_n;\A_{-n}^{(t)})+h_n(\A_n)$, inexact BCD updates $\A_n$ via minimizing a local approximation of $f(\A_n;\A_{-n}^{(t)})+h_n(\A_n)$ at $\A_n=\A_n^{(t)}$ which we denote 
 $$
 g(\A_n;{\cal A}^{(t)}) \approx f(\A_n;\A_{-n}^{(t)})+h_n(\A_n),
 $$
where 
${\cal A}^{(t)}=\{ \A_1^{(t+1)},\ldots,\A_{n-1}^{(t+1)},\A_n^{(t)},\ldots,\A_N^{(t)} \}$.
That is, inexact BCD updates $\A_n$ using 
$$
\A_n^{(t+1)}\leftarrow  \arg\min_{\A_n}~g(\A_n;{\cal A}^{(t)}).
$$						
If $g(\A_n;{\cal A}^{(t)})$ admits a simple minimizer, then the algorithm can quickly update $\A_n$ and move to the next block.
One of the frequently used  $g(\A_n;{\cal A}^{(t)})$ is as follows:
\begin{align}\label{eq:g}
 g(\A_n;{\cal A}^{(t)}) &= f(\A_n^{(t)};\A_{-n}^{(t)})+h_n(\A_n)\\
 +\nabla_{\A_n} f(\A_n^{(t)};\A_{-n}^{(t)})^\T&(\A_n-\A_n^{(t)}) + \frac{1}{2\alpha}\|\A_n-\A_n^{(t)}\|_{\rm F}^2, \nonumber
\end{align}
which is obtained via applying the Taylor's expansion on the smooth term $f(\bm \theta)$.
Using the above local approximation, the update admits the following form:
\begin{equation}\label{eq:inexactPGD}
\A_n^{(t+1)}\leftarrow  {\sf Prox}_{h_n}\left( \A_n^{(t)}  - \alpha \nabla_{\A_n} f\left(\A_n^{(t)};\A_{-n}^{(t)} \right) \right),
\end{equation}
which is equivalent to running PGD for one iteration to solve \eqref{eq:cls}---and this echoes the term ``inexact''. A number of frequently used local approximations can be seen in \cite{razaviyayn2013unified}.
Note that inexact BCD is not unfamiliar to the SLRD community, especially for nonnegativity constraints. One of the most important early algorithms for NMF, namely, the multiplicative updates (MU) \cite{seung2001algorithms}, is an inexact BCD algorithm. 

\subsection{Pragmatic Acceleration} \label{sec:pragmaaccel}
\begin{figure}
	\centering
	\setlength{\figurewidth}{5cm}
    \setlength{\figureheight}{2.6cm}
    \input{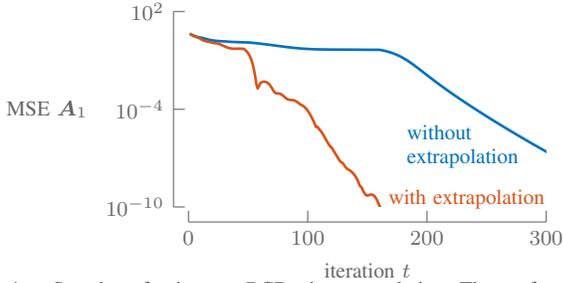}
    \vspace{-5mm}
	\caption{Speed-up for inexact BCD via extrapolation. The performance is measured by the mean squared error (MSE) on the estimated latent factor; see~\cite{fu2019block}. The tensor size is $30\times 30\times 30\times 30$ and $R=10$. The inexact BCD and extrapolated version use the updates in \eqref{eq:inexactPGD} and \eqref{eq:extra} \cite{xu2013block}, respectively.} 
	\vspace{-.5cm}
	\label{fig:fu2016robust}
\end{figure}

Compared to exact BCD, inexact BCD normally needs to update all block variables many more rounds before reaching a ``good'' solution. Nonetheless, when inexact BCD is combined with the so-called ``extrapolation'' technique, the convergence speed can be substantially improved. The procedure of extrapolation is as follows: Consider an extrapolated point
\begin{equation}\label{eq:extra}
       \widehat{\A}_n^{(t)} =(1+ \omega_n^t) \A_n^{(t)} + \omega_n^t \A_n^{(t-1)},  
\end{equation}
where $\{\omega_n^t\}$ is a pre-defined sequence (see practical choices of $\omega_n^t$ in \cite{xu2013block}).
Then, the extrapolation-based inexact BCD replaces \eqref{eq:inexactPGD} by the following:
$$
\A_n^{(t+1)}\leftarrow  {\sf Prox}_{h_n}\left( \widehat{\A}_n^{(t)}  - \alpha \nabla_{\A_n} f\left(\widehat{\A}_n^{(t)};\A_{-n}^{(t)} \right) \right).
$$ 
In practice, this simple technique oftentimes makes a big difference in terms of convergence speed; see Fig.~\ref{fig:fu2016robust}.
The extrapolation technique was introduced by Nesterov in 1983 to accelerate smooth single-block convex optimization problems using only first-order derivative information { \cite{nesterov1983method}.}
It was introduced to handle nonconvex, multi-block, and nonsmooth problems in the context of tensor decomposition by Xu \textit{et al.} in 2013 \cite{xu2013block}. In this case, no provable acceleration has been shown, which leaves a challenging and interesting research question open.

\subsection{Convergence Properties and Computational Complexity}

Convergence properties of both exact BCD and inexact BCD are well studied in the literature \cite{bertsekas1999nonlinear,razaviyayn2013unified}.
An early result from Bertsekas \cite{bertsekas1999nonlinear} shows that every limit point of $\{ \bm \theta^{(t)} \}_t$ is a stationary point of Problem~\eqref{eq:problem}, if $F$ is absent or is the indicator function of a convex closed set, and if the subproblems in \eqref{eq:bcd} can be exactly solved with unique minimizers while the objective function is non-increasing in the interval between two consecutive iterates. 
This can be achieved if the subproblems in \eqref{eq:bcd} are strictly (quasi-)convex.
Nonetheless, since $\H_n^{(t)}$ may be rank deficient, stationary-point convergence under this framework is not necessarily easy to ensure. In addition, this early result does not cover nonsmooth functions.

For inexact BCD, it was shown in \cite{razaviyayn2013unified} that if $h_n( \A_n)$ is convex, and if 
the local surrogate is strictly (quasi-)convex and is ``tangent'' to $F(\A_n;\A_{-n}^{(t)})$ at $\A_n=\A_n^{(t)}$ (i.e., it is tight and shares the same directional derivatives at this point), then every limit point of the produced solution sequence is a stationary point. 
This is a somewhat more relaxed condition relative to those for BCD, since the upper bound $g(\A_n;{\cal A}^{(t)})$ can always be constructed as a strictly convex function, e.g., by using \eqref{eq:g}.

In terms of per-iteration complexity, BCD combined with first-order subproblem solvers for structured tensor decomposition is not a lot more expensive than solving unconstrained ones in many cases---which is the upshot of using algorithms like PGD, accelerated PGD, or ADMM.
The most expensive operation is the so-called matricized tensor times Khatri--Rao product (MTTKRP), i.e., $\X_n^\T\H_{n}^{(t)}$. However, even if one uses exact BCD with multiple iterations of PGD and ADMM for solving \eqref{eq:cls}, the MTTKRP only needs to be computed once for every update of $\A_n$, which is the same as in the unconstrained case; see more discussions in \cite{huang2016flexible}.

\subsection{Block Splitting and Ordering Within BCD}\label{sec:blocksplitting}

In this paper, we focus on the most natural choice of blocks to perform BCD on low-rank matrix/tensor decomposition models, namely, $\{\A_n\}_{n=1}^N$.  
However, in some cases, it might be preferable to optimize over smaller blocks because the subproblems are simpler. For example, with nonnegativity constraints, it has been shown that optimizing over the blocks made of the columns of the $\A_n$'s is rather efficient (because there is a closed-form solution) and outperforms exact BCD and the MU~\cite{gillis2014and, zhou2014nonnegative} that are based on $\A_n$-block splitting. 
Another way to modify BCD and possibly improve convergence rates is to update the blocks of variables in a non-cyclic way; 
for example, using random shuffling at each outer iteration, or picking the block of variables to update using some criterion that increases our chances to converge faster 
(e.g., pick the block that was modified the most in the previous iteration, i.e., pick $\argmin_n \| \A_n^{(t)} - \A_n^{(t-1)} \| / \|\A_n^{(t-1)}\|_\text{F}$); 
see, e.g., \cite{hsieh2011fast,li2015convergence}.

\section{Second-order Approaches}
\label{sec:org1d8a73b}

\begin{figure}[htp]
  \setlength{\figurewidth}{5cm}
  \setlength{\figureheight}{3cm}
  \centering
  \input{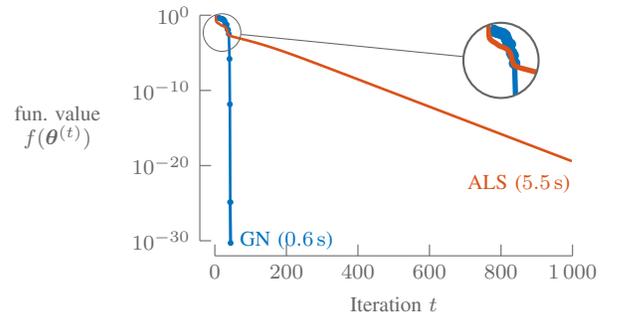}
  \caption{While ALS initially improves the function value faster, the
    Gauss--Newton (GN) method converges more quickly, both in terms of time and
    number of iterations. Results shown for a $100\times100\times100$ rank $R=10$ tensor
    with highly correlated rank-1 terms (average angle is \SI{59}{\degree}),
    starting from a random initialization.}
  \label{fig:convergence-als-nls}
\end{figure}

Combining SLRD and optimization techniques that exploit (approximate)
second-order information has a number of advantages. Empirically, these
algorithms converge in much fewer iterations relative to first-order methods,
are less susceptible to the so-called swamps, and are often more robust to
initializations
\cite{sorber2013optimizationbasedalgorithms,phan2013lowcomplexity}; 
see, e.g., Fig.~\ref{fig:convergence-als-nls}.

There are many second-order optimization algorithms, e.g., the Newton's method
that uses the Hessian and a series of ``quasi-Newton''
methods that approximate the Hessian.  Among these algorithms, the Gauss--Newton
(GN) framework specialized for handling nonlinear least squares (NLS) problems fits
Euclidean distance based tensor/matrix decompositions particularly well.  Under
the GN framework, the structure inherent to some tensor models (e.g., CPD and
LL1) can be exploited to make the per-iteration complexity of the same order
as the first-order methods \cite{sorber2013optimizationbasedalgorithms}.

\subsection{Gauss--Newton Preliminaries}\label{sec:org58f0563}
Consider the unconstrained tensor decomposition problem:
\begin{align}
  \min_{\bm{\theta}} f(\bm{\theta}) \text{\ \ with\ \ } f(\bm{\theta}) = \frac12
  \|\underbrace{\cp{\A_1,\ldots,\A_N}-\ten{T}}_{\ten{F}(\bm \theta)}\|^2_{\rm F}.
  \label{eq:nls-obj}
\end{align}
The GN method starts from a linearization of the residual \(\ten{F}\):
\begin{align}
  \vectorize{\ten{F}(\bm{\theta})}
  &\approx \vectorize{\ten{F}(\bm{\theta}^{(t)})} +
  \left.\frac{\text{d}\,\vectorize{\ten{F}}}{\text{d}\,\vectorize{\bm{\theta}}}
    \right|_{\bm{\theta}^{(t)}}\cdot\bm{p}\\
  &= \vec{f}^{(t)} + \bm{J}^{(t)} \bm{p},\label{eq:nls-taylor}
\end{align}
where \(\bm{J}^{(t)}\) is the Jacobian of \(\ten{F}\) w.r.t.~the variables
\(\bm{\theta}\), $\bm p=\bm{\theta} - \bm{\theta}^{(t)}$, and \(\bm{f}^{(t)} =
\vectorize{\ten{F}(\bm{\theta}^{(t)})}\). Substituting \eqref{eq:nls-taylor} in \eqref{eq:nls-obj} results in a quadratic optimization problem
\begin{align}
  \bm p^{(t)}&\leftarrow \argmin_{\bm{p}}\ \frac12
  \|\bm{f}^{(t)}\|^2 + \vec{g}^{(t)^{\T}} \bm{p}
               + \frac12\bm{p}^{\T} \bm{\varPhi}^{(t)}
               \bm{p},\label{eq:gn-obj}
\end{align}
in which the gradient is given by $\bm g^{(t)}=\bm J^{(t)^\T} \bm f^{(t)}$ and
the Gramian (of the Jacobian) by
$\bm \varPhi^{(t)}=\bm{J}^{(t)^{\T}}\bm{J}^{(t)}$. The variables are updated as
$ \bm \theta^{(t+1)} \leftarrow \bm \theta^{(t)} + \bm p^{(t)}$. We have
\begin{align}
  \label{eq:system-nls}
  \bm \varPhi^{(t)} \bm p^{(t)} =  -\bm g^{(t)},
\end{align}
by the optimality condition of the quadratic problem in \eqref{eq:gn-obj}.
In the case of CPD, the Gramian \(\bm{J}^{(t)^{\T}}\bm{J}^{(t)}\) is a positive
semidefinite matrix instead of a positive definite { one}\footnote{The Gramian
\(\bm{J}^{(t)^{\T}}\bm{J}^{(t)}\) has at least \((N-1)R\) zero eigenvalues because of the
scaling indeterminacy.}, which means that \(\bm{p}^{(t)}\) is not an ascent
direction, but may not be a descent direction either. This problem can be
avoided by the Levenberg--Marquardt (LM) method, i.e., using $\bm{\varPhi}^{(t)} = \bm{J}^{(t)^{\T}}\bm{J}^{(t)} + \lambda
\bm{I}$ for some \(\lambda \geq 0\), or using a trust region which implicitly
dampens the system. The GN method can exhibit up to quadratic convergence rate near an optimum if the
residual is small \cite{sorber2013optimizationbasedalgorithms,bertsekas1999nonlinear}.

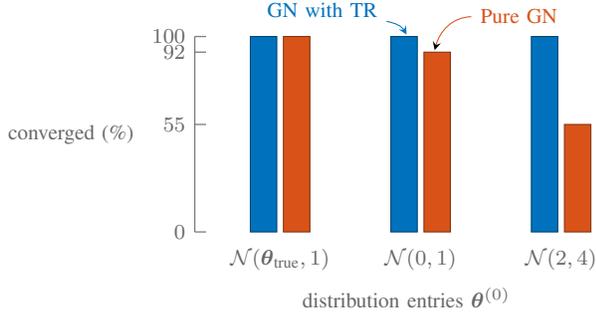
\begin{figure}[tp]
  \centering
  \definecolor{mycolor1}{rgb}{0.00000,0.44700,0.74100}%
  \definecolor{mycolor2}{rgb}{0.85000,0.32500,0.09800}%
  \setlength{\figurewidth}{5.5cm}
  \setlength{\figureheight}{2.6cm}
\begin{tikzpicture}[font=\footnotesize]
\begin{axis}[%
width=0.95092\figurewidth,
height=\figureheight,
at={(0\figurewidth,0\figureheight)},
scale only axis,
area legend,
xmin=0.5,
xmax=3.3,
xtick={1,2,3},
xticklabels={{$\ten{N}(\vec{\theta}_{\text{true}},1)$}, {$\ten{N}(0,1)$}, {$\ten{N}(2,4)$}},
xlabel={distribution entries $\vec{\theta}^{(0)}$},
ylabel={converged (\%)},
ytick={0,55,92,100},
ymin=0,
ymax=100,
minimal x axis offset = -1mm,
minimal,
every plot/.append style={ybar,bar width=0.172895\figurewidth,draw=black},
every axis/.append style={%
  every outer x axis line/.append style = { draw = white },
},
every x tick/.append style = { line width = 0.5pt, draw=white },
ylabel style={rotate=-90,text width=2.0cm,align=right},
]
\addplot[ybar,draw=mycolor1!60!black,fill=mycolor1,bar shift=-0.04\figurewidth] plot table[row sep=crcr] {%
1	100\\
2	100\\
3	100\\
};
\addplot[ybar,draw=mycolor2!60!black,fill=mycolor2,bar shift=0.04\figurewidth] plot table[row sep=crcr] {%
1	100\\
2	92\\
3	55\\
};

\node at (axis cs:1.3,105) [text=mycolor1,anchor=south] (a) {GN with TR};
\node at (axis cs:2.7,102) [text=mycolor2,anchor=south] (b) {Pure GN};

\draw [->,>=stealth,draw=mycolor1] (a) to [out=-10,in=120] (axis cs:1.9,101);
\draw [->,>=stealth,draw=mycolor2] (b) to [out=181,in=75] (axis cs:2.1,93);
 
\end{axis}
\end{tikzpicture}%
  
\caption{Without globalization strategy, pure Gauss--Newton (GN) does not always
  converge to a stationary point; by using a dogleg trust region (TR) based globalization, GN converges for every
  initialization $\vec{\theta}^{(0)}$. Results shown for a $20\times20\times20$ rank-10 tensor in
  which all factor matrix entries $\vec{\theta}_{\text{true}}$ are draw from the
  normal distribution $\ten{N}(0,1)$, and 100 different initializations
  for three scenarios ranging from good (left) to bad (right).}
  \label{fig:globalization}
\end{figure}

Second-order methods converge fast once \(\bm{\theta}^{(t)}\) is near a stationary point, while there is a risk that $\bm \theta^{(t)}$ may never come close to any stationary point.
To ensure global convergence, i.e., that \(\bm{\theta}^{(t)}\) converges to a
stationary point from any starting point \(\bm{\theta}^{(0)}\), {\it globalization} strategies can be used
\cite{bertsekas1999nonlinear}. 
Globalization is considered crucial for nonconvex optimization based tensor decomposition algorithms and makes them robust w.r.t. the initial guess $\vec{\theta}^{(0)}$, as is illustrated in Fig.~\ref{fig:globalization}.

The first effective globalization strategy is determining \(\alpha^{(t)}\) via solving the following:
\begin{align}
  \alpha^{(t)} = \argmin_{\alpha} f(\bm{\theta}^{(t)}+\alpha\bm{p}^{(t)}), \label{eq:linesearch}
\end{align}
which is often referred to as \emph{exact line search} in the optimization literature.
Solving the above can be costly in general, but when the objective is to
minimize a multilinear error term in least squares sense as in~\eqref{eq:nls-obj}, the global minimum of this problem can be found exactly,
as the optimality conditions boil down to a polynomial root finding problem; see
\cite{sorber2015exactline} and references therein. This exact line search
technique ensures that the maximal progress is made in every step of GN, which helps improve the objective function quickly. Similarly, \emph{exact plane search}
  can be used to find the best descent direction in the plane spanned by $\bm p^{(t)}$ and $\bm g^{(t)}$ by searching for coefficients $\alpha$ and $\beta$ that minimize $f(\vec{\theta}^{(t)}+\alpha \vec{p}^{(t)} + \beta \vec{g}^{(t)})$
  \cite{sorber2015exactline}. Empirically, the steepest descent
  direction $-\bm g^{(t)}$ decreases the objective function more rapidly during
  earlier iterations, while the GN step allows fast convergence. Note that plane search can be used to speed up BCD techniques as well
  \cite{sorber2015exactline}.

Another effective globalization strategy uses a trust region (TR). There, the problems of finding the step direction
\(\bm{p}^{(t)}\) and step size are combined, i.e., \(\bm{\theta}^{(t+1)} =
\bm{\theta}^{(t)} + \bm{p}^{(t)}\) with
\begin{align}
  \bm{p}^{(t)} = \argmin_{\bm{p}} m(\bm{p}) \quad \text{s.t.}
  \quad \norm{\bm{p}} \leq \Delta,\label{eq:trustregion}
\end{align}
where $m(\bm p)=\frac12 \|\bm{f}^{(t)}\|^2
+ \bm{g}^{(t)^{\T}}\bm{p}
+ \frac12\bm{p}^{\T} \bm{\varPhi}^{(t)} \bm{p}$ under the GN framework.
Intuitively, the TR is employed to prevent the GN steps to be too aggressive to miss the contraction region of a stationary point.
The TR radius \(\Delta\) is determined
heuristically by measuring how well the model predicts the decrease in function
value
\cite{bertsekas1999nonlinear}. 
Often,
the search space is restricted to a two-dimensional subspace spanned by
\(\bm{g}^{(t)}\) and \(\bm{p}^{(t)}\). Problem~\eqref{eq:trustregion} can then be solved
approximately using the dogleg step, or using plane search
\cite{sorber2015exactline,sorber2013optimizationbasedalgorithms}. 


\subsection{Exploiting Tensor Structure}
\label{sec:orgea1a010}

The bottleneck operation in the GN approach is constructing and solving the
linear system in~\eqref{eq:system-nls}, i.e.,
\begin{align}
\bm{J}^{\T}\bm{J} \bm{p} = -\bm{g} \label{eq:system-gn}, 
\end{align}
where the superscripts \((\cdot)^{(t)}\) have been dropped for simplicity of
notation. Note that this system is easily large scale, since
$\bm J\in\mathbb{R}^{\prod_nI_n\times T}$ where $T=R(\sum_{n=1}^I I_n)$.
Using a general-purpose solver for this system costs ${\cal O}(T^3)$ flop, which may be prohibitive for big data problems.
Fortunately, the Jacobian and the Gramian are both structured under certain
decomposition models (e.g., CPD and LL1), which can be exploited to come up
with lightweight solutions.

The gradient $\vec{g}=\nabla f(\bm \theta)$ of
\(f(\bm{\theta})\) can be partitioned as
\(\bm{g} = [\vectorize{\bm{G}_{1}}; \ldots; \vectorize{\bm{G}_N}]\), in which
the \(\bm{G}_{n}\) w.r.t.~factor matrix \(\A_{n}\) is given by
\begin{align}
  \bm{G}_{n} &= \bm{F}_{n}^\T \bm{H}_{n},
\end{align}
in which $\mat{F}_n$ is the unfolding of the residual $\ten{F}$; 
see~\eqref{eq:matrix-unfolding}. 
The operation $\bm{F}_{n}^\T \bm{H}_{n}$ is the
well-known MTTKRP as we have seen in the BCD approaches.  However, the factor
matrices \(\A_{n}\) (\(n=1,\ldots,N\)) have the same value for every gradient
\(\bm{G}_{n}\), in contrast to BCD algorithms which uses updated variables in
every inner iteration. This can be exploited to reduce the computational cost
\cite{phan2013fastalternating}.

Similarly to the gradient, the Jacobian \(\bm{J}\) can also be partitioned as \(\bm{J} = [\bm{J}_{1}, \ldots,
\bm{J}_{N}]\) in which \(\bm{J}_{n} =
\frac{\partial\vectorize{\ten{F}}}{\partial \vectorize{\A_{n}}}\):
\begin{align}
  \bm{J}_{n} = \bm{\Pi}_{n} (\bm{H}_n\kron \bm{I}_{I_n}),\label{eq:jacobian}
\end{align}
in which \(\bm{\Pi}_{n}\) is a matrix corresponding to permutation of mode 1 to
mode \(n\) of vectorized tensors. By exploiting the block and Kronecker
structures, constructing \(\bm{\varPhi}\) requires only \(\order{T^2}\)
flop, as opposed to ${\cal O}(T^3)$; for details, see  \cite{phan2013lowcomplexity,vervliet2018numericaloptimization}.

Instead of solving~\eqref{eq:system-gn} exactly, an iterative solver such as conjugate gradients (CG) can
be used. As in power iterations, the key step in a CG iteration is a
Gramian-vector product, i.e., given $\vec{\upsilon}$ compute $\vec{y}$ as:
\begin{align}
  \bm{y} = \bm{J}^{\T}\bm{J}\bm{\upsilon}.  \label{eq:gramvec-prod}
\end{align}
Both \(\bm{y}\) and \(\bm{\upsilon}\) can be partitioned according to the variables,
hence \(\bm{\upsilon} = [\bm{\upsilon}_{1};\ldots;\bm{\upsilon}_{N}]\) and \(\bm{y} =
[\bm{y}_{1};\ldots;\bm{y}_{N}]\). 
Eq.~\eqref{eq:gramvec-prod} can then be
written as 
$\bm{y}_{n}= \bm{J}_{n}^{\T}\sum_{k=1}^{N}\bm{J}_{k}\bm{\upsilon}_{k} 
$, which is computed efficiently by exploiting the structure in \(\bm{J}_{n}\)
[cf.\ Eq.~\eqref{eq:jacobian}]: 
\begin{align}
  \bm{Y}_{n} &= \bm{V}_{n}\bm{W}_n +
                  \A_{n}\sum_{\substack{k=1\\k\neq n}}^{N}
  \bm{W}_{kn} \hadamard (\bm{V}_{k}^{\T}\A_{k}), \label{eq:gn-matvec}
\end{align}
where $\mat{Y}_n,\mat{V}_n\in\R^{I_n\times R}$ and $\vec{y}_n = \vectorize{\bm Y_n}$ and $\vec{\upsilon}_n = \vectorize{\bm V_n}$, resp.
${\bm W}_n$ and $\bm W_{kn}$ are defined as follows:
\begin{align}\label{eq:def_w}
	\bm{W}_n &= \hadamard_{\substack{k=1\\k\neq n}}^{N}
	\A_{k}^{\T}\A_{k}, &
	\bm{W}_{mn} &= \hadamard_{\substack{k=1\\k\neq m,n}}^{N}
	\A_{k}^{\T}\A_{k}.
\end{align}
Hence, to compute \(\bm{J}^{\T}\bm{J}\bm \upsilon\) only products of small \(I_n\times R\) and
\(R\times R\) matrices are required. As CG performs a number of iterations with
constant \(\A_{n}\), the inner products \(\A_{n}^{\T}\A_{n}\) required for
\(\bm{W}_n\) and \(\bm{W}_{kn}\) can be precomputed. This way, the complexity per
Gramian-vector product is only \(\order{R^2\sum_{n}I_n}\). Note that for both GN
and the BCD methods, the computation of the gradient---which requires
$\order{R\prod_{n=1}^{N}I_n}$ operations---usually dominates the
complexity. Therefore, the GN approach is also an excellent candidate for
parallel implementations as it reduces the number of iterations and expensive
gradient computations, while the extra CG iterations have a negligible
communication overhead. 

In practice, it is important to notice that a well-conditioned $\bm J^\T\bm J$
makes solving the system in \eqref{eq:system-gn} much faster using CG.
In numerical linear algebra, the common practice is to precondition
$\bm J^\T\bm J$, leading to the so-called preconditioned CG (PCG)
paradigm. Preconditioning can be done rather efficiently under some LRDMs like CPD; see the insert ``{\bf Acceleration via Preconditioning}''.

\label{sec:org0155e50}
\begin{mdframed}[backgroundcolor=blue!10,topline=false,
	rightline=false,
	leftline=false,
	bottomline=false]
\noindent
{\bf Acceleration via Preconditioning}
As the convergence speed of CG depends on the `clustering of the
eigenvalues' of \(\bm{J}^{\T}\bm{J}\), a
preconditioner is often applied to improve this clustering. More specifically,
the system 
\begin{align}
  \bm{M}^{-1}\bm{J}^{\T}\bm{J}\bm{p} = -\bm{M}^{-1}\bm{g}
  \label{eq:gn-system-prec}
\end{align}
is solved instead of~\eqref{eq:system-gn} and the preconditioner \(\bm{M}\) is
chosen to reduce the computational cost.
In practice, a Jacobi preconditioner, i.e., \(\bm{M}\) is a diagonal
matrix with entries \(\diag(\bm{J}^{\T}\bm{J})\), or a block-Jacobi
preconditioner, i.e., a block-diagonal approximation to \(\bm{J}^{\T}\bm{J}\),
are often effective for the unconstrained CPD
\cite{sorber2013optimizationbasedalgorithms}. For example, the latter
preconditioner is given by
\begin{align}
  \bm{M}_{\text{BJ}}
  &= \text{blkdiag}(\bm{W}_1\kron
     \bm{I}_{I_1},\ldots,\bm{W}_N\kron \bm{I}_{I_N}).
\end{align}
Because of the block-diagonal and Kronecker structure in \(\bm{M}_{\text{BJ}}\),
the system \(\bm{\upsilon} = \bm{M}^{-1}_{\text{BJ}} \bm{y}\) can be solved in $N$
steps, i.e., $\bm{V}_{n} = \bm{Y}_{n} \bm{W}_n^{-1}$ for $n=1,\ldots,N$.
Applying
$\mat{M}_{\text{BJ}}^{-1}$ only involves inverses of small \(R\times R\)
matrices which are constant in one GN iteration. Interestingly,
$\mat{M}_{\text{BJ}}$ appears in the ALS algorithm with simultaneous
updates---i.e., without updating $\mat{A}_n$ every inner iteration. The PCG
algorithm can therefore be seen as a refinement of ALS with simultaneous updates
by taking the off-diagonal blocks into account
\cite{sorber2013optimizationbasedalgorithms}.
\end{mdframed}

\subsection{Structured Decomposition}
\label{sec:org1dab277}
As mentioned, the GN framework is specialized for NLS problems, i.e., objectives
can be written as $\|{\cal F}(\bm \theta)\|_{\text{F}}^2$. If there are structural
constraints on $\bm \theta$, incorporating such structural requirements is often
nontrivial. In this subsection, we introduce a number of ideas for handling
structural constraints under the GN framework.

\subsubsection*{Parametric constraints}
\label{sec:org36bab46}
One way to handle constraints is to use parametrization to convert the constrained decomposition problem to an unconstrained NLS problem. To see how it works, let us consider the case where $h_n(\bm \theta_n)$ is an indicator function of set ${\cal C}_n$, i.e., the constrained decomposition case where $\bm \theta_n \in {\cal C}_n$. 
In addition, we assume that every element in ${\cal C}_n$ can be parameterized by an unconstrained variable.
Assume $\bm \theta_n={\rm vec}(\A_n)$ and every factor matrix
\(\A_{n}\) is a function \(q_{n}\) of a disjoint set of parameters
\(\bm{\alpha}_{n}\), i.e., \(\A_{n} = q_{n}(\bm{\alpha}_{n})\),
\(n=1,\ldots,N\). For example, if ${\cal C}_n=\mathbb{R}^{I_n\times R}_{+}$, i.e., the nonnegative orthant,
one can parameterize $\A_n$ using the following:
\[    \A_n=\bm D \circledast \bm D,~\bm D\in\mathbb{R}^{I_n\times R}.\]
In this case, $\bm \alpha_n={\rm vec}(\bm D)$ and $q_n(\cdot):\mathbb{R}^{I_nR}\rightarrow \mathbb{R}^{I_nR}$ denotes the elementwise squaring operation.
If no constraint is imposed on some \(\A_{n}\), \(q_{n}(\bm{\alpha}_{n}) = \text{unvec}(\bm{\alpha}_{n})\); see many other examples for different constraints in \cite{vervliet2018numericaloptimization}. 

By substituting the constraints in the optimization problem \eqref{eq:nls-obj}, we
obtain a problem in variables \(\bm{\alpha} = [\bm{\alpha}_{1}; \ldots; 
\bm{\alpha}_{N}]\):
\begin{align}
  \min_{\bm{\alpha}} \frac12 \norm{\ten{T} -
  \cp{q_{1}(\bm{\alpha}_{1}),\ldots,q_{N}(\bm{\alpha}_{N})}}^2_{\text{F}}. 
  \label{eq:obj-gn-constrained}
\end{align}
Applying GN to~\eqref{eq:obj-gn-constrained} follows the same steps as before. Central to this unconstrained problem is the solution of
\begin{align}
  \tilde{\bm{\varPhi}} \tilde{\bm{p}} = -\tilde{\bm{g}},  \label{eq:nls-constrained-system}
\end{align}
where we denote quantities related to parameters by tildes to distinguish them from
quantities relates to factor matrices. The structure of the factorization models can
still be exploited if we use the chain rule for derivation \cite{sorber2015structureddatafusion}. This way, \eqref{eq:nls-constrained-system} can be written as
\begin{align}
  \tilde{\bm{J}}^{\T} \bm{\varPhi} \tilde{\bm{J}} \tilde{\bm{p}} =
  -\tilde{\bm{J}}^{\T}\bm{g}, \label{eq:nls-constrained-system-expanded}
\end{align}
in which \(\bm{\varPhi}\) and \(\bm{g}\) are exactly the expressions as derived
before in the unconstrained case. The Jacobian \(\tilde{\bm{J}}\) is a block diagonal matrix containing
the Jacobian of each factor matrix w.r.t.~the underlying variables, i.e.,
\begin{align}
  \tilde{\bm{J}} =
  \text{blkdiag}(\tilde{\bm{J}}_{1},\ldots,\tilde{\bm{J}}_{N}).
\end{align}
The Jacobians \(\tilde{\bm{J}}_{n}\) are often straightforward to derive. For example,
if \(\A_{n}\) is unconstrained, \(\tilde{\bm{J}}_{n} = \bm{I}_{I_nR}\);
if nonnegativity is imposed by squaring variables, \(\A_{n} = \bm{D} \circledast
\bm{D}\) and \(\tilde{\bm{J}}_{n} = \diag(\vectorize{2\bm{D}})\); in the case of linear
constraints, e.g., \(\A_{n} = \bm{B}\bm{X}\bm{C}\) with
\(\bm{B}\) and \(\bm{C}\) known, \(\tilde{\bm{J}}_{n} = \bm{C}^{\T}\kron
\bm{B}\). More complicated constraints can be modeled via composite functions and by applying the chain rule repeatedly
~\cite{vervliet2018numericaloptimization,sorber2015structureddatafusion}. 

When computing the Gramian or Gramian vector products in~\eqref{eq:nls-constrained-system-expanded}, we can exploit the multilinear
structure from the CPD as well as the block-diagonal structure of the
constraints. Moreover, depending on the constraint, $\tilde{\mat{J}}_n$ may, for
example, also have diagonal or Kronecker product structure. Therefore, the
Gramian-vector products can be computed in three steps:
\begin{align}
  \bm{\upsilon}_{n} = \tilde{\bm{J}}_{n} \tilde{\bm{\upsilon}}_{n}, \quad \bm{y}_{n} = \bm{J}_{n}^{\T}\sum_{k=1}^{N} \bm{J}_{k}\bm{\upsilon}_{k},\quad
  \tilde{\bm{y}}_{n} = \tilde{\bm{J}}_{n}^{\T} \bm{y}_{n},\label{eq:nls-constrained-contract}
\end{align}
which may all be computed efficiently using similar ideas as in the unconstrained case [cf.\ Eq.~\eqref{eq:gn-matvec}].
Leveraging the chain rule and the Gramian-vector product based PCG method for handling the unconstrained GN framework, it turns out that many frequently used constraints in signal processing and data analytics can be handled under this framework in an efficient way. Examples include nonnegativity, polynomial constraints, orthogonality, matrix inverses, Vandermonde, Toeplitz or Hankel structure; see details in \cite{vervliet2018numericaloptimization}.

We should mention that the parametrization technique can also handle some special constraints that are considered quite challenging in the context of tensor and matrix factorization, e.g., (partial) symmetry and coupling constraints; see the insert in ``{\bf Handling Special Constraints via Parametrization}''.

\begin{mdframed}[backgroundcolor=blue!10,topline=false,
	rightline=false,
	leftline=false,
	bottomline=false]

\noindent
{\bf Handling Special Constraints via Parameterization}
Factorizations are often (partially) symmetric, e.g., in blind source
separation and topic modeling (see examples in the tutorial \cite{sidiropoulos2017tensor}). Symmetry
here means that some $\A_n$'s are
identical. 
The conventional BCD treating each $\A_n$ as a block is not straightforward
anymore. For example, cyclically updating the factor matrices $\mat{A}_1
  = \mat{A}_2 = \mat{A}_3 = \mat{A}$ in the
  decomposition $\cp{\mat{A},\mat{A},\mat{A}}$ breaks symmetry, while the
  subproblems are no longer convex when enforcing symmetry; see also
Sec.~\ref{sec:blocksplitting}. Nevertheless, the GN
framework handles such constraints rather naturally.

A (partially) symmetric CPD can be modeled by setting two or more factors to be
identical. For example, consider the model
\(\cp{\mat{A}_1,\ldots,\mat{A}_{N-2},\mat{A}_{N-1},\mat{A}_{N-1}}\), i.e., the
last two factor matrices are identical. This symmetry constraint leads to a \(\tilde{\mat{J}}\) with the following form:
\begin{align*}
  \tilde{\mat{J}} =
  \text{blkdiag}(\mat{I}_{I_1R},\ldots,\mat{I}_{I_{N-2}R},[\mat{I}_{I_{N-1}R};
  \mat{I}_{I_{N-1}R}]),
\end{align*} 
in which \(\tilde{\mat{J}}_{n}\), \(n=1,\ldots,N-1\), are identity matrices as no
constraints are imposed on \(\mat{A}_n\). Because of the structure in $\tilde{\mat{J}}$, the extra steps in the Gramian-vector products in \eqref{eq:gramvec-prod} only involve summations.

Coupled decomposition often arises in data fusion, e.g., integrating hyperspectral and multispectral images for super-resolution purposes \cite{wei2015fast}, { spectrum cartography from multiple sensor-acquired spatio-spectral information \cite{zhang2020spectrum}}, or jointly analyzing primary data and side information \cite{beutel2014flexifact}. These problems involve jointly factorizing tensors and/or matrices: the decompositions can share,
or are coupled through, one or more factors or underlying variables. For
example, consider the coupled matrix tensor factorization problem:
\begin{equation*}
   \min_{\{\mat{A}_n\}_{n=1}^4} \hspace{-0.15cm} 
    \frac{\lambda_1}{2} \hspace{-0.05cm}  \norm{\cp{\mat{A}_1,\mat{A}_2,\mat{A}_3}\hspace{-0.05cm}\text{$-$}\ten{T}}^2_{\text{F}}\text{$+$}\frac{\lambda_2}{2} \hspace{-0.05cm} \norm{\cp{\mat{A}_3,\mat{A}_4}\hspace{-0.05cm}\text{$-$}\mat{M}}^2_{\text{F}} 
  \label{eq:nls-obj-coupled-example}
\end{equation*}
where the two terms are coupled through \(\mat{A}_3\). 
Coupled decomposition can be handled via BCD. However, in some cases, the key steps of BCD boil down to solving Sylvester equations in each iteration, which can be costly for large-scale problems \cite{wei2015fast}. Using parametrization and GN, the influence of the coupling constraint and the decomposition can be separated in the CG iterations  \cite{vervliet2018numericaloptimization,sorber2015structureddatafusion}---and thus easily puts forth efficient and flexible data fusion algorithms. This serves as the foundation of the structured data fusion (SDF) toolbox in \texttt{Tensorlab}.

\end{mdframed}

\subsubsection*{Proximal Gauss--Newton}
To handle more constraints and the general cost function $f(\bm \theta)+h(\bm \theta)$ in a systematic way, one may also employ the proximal GN (ProxGN) approach.
To be specific, in the presence of a nonsmooth $h(\bm \theta)$, the ProxGN framework modifies the per-iteration sub-problem of GN into 
\begin{align}
  \bm \theta^{(t+1)}\leftarrow \arg\min_{\bm \theta}\frac12
  \|\bm f^{(t)} + \bm{J}^{(t)}(\bm{\theta}-\bm{\theta}^{(t)})\|_2^2+h(\bm \theta).
  \label{eq:pgn-obj}
\end{align}
This is conceptually similar to the PGD approach: linearizing the smooth part (using the same linearization as in unconstrained GN) while keeping the nonsmooth regularization term untouched.
The subproblem in \eqref{eq:pgn-obj} is again a regularized least squares problem w.r.t. $\bm \theta$.
Similar to the BCD case [cf. Eq.~\eqref{eq:cls}], there exists no closed-form solution for the sub-problem in general.
However, subproblem solvers such as PGD and ADMM can again be employed to handle the \eqref{eq:pgn-obj}.

A recent theoretical study has shown that incorporating the proximal term does
not affect the overall super-linear convergence rate of the GN-type algorithms
within the vicinity of the solution. The challenge, however, is to solve \eqref{eq:pgn-obj} in the context SLRD with lightweight updates.  This is possible. The recent paper
in \cite{huang2019globalsip} has shown that if ADMM is employed, then the key
steps for solving \eqref{eq:pgn-obj} are essentially the same as that of the
unconstrained GN, namely,
computing
$(\bm{J}^{(t)\T}\bm{J}^{(t)}+\rho\bm{I})^{-1}$ for a certain $\rho>0$ once per
ProxGN iteration.  Note that this step is nothing but inverting the regularized
Jacobian Gramian, which, as we have seen, admits a number of economical solutions. In addition, with judiciously designed ADMM steps, this Gramian inversion never needs to be instantiated---the algorithm is memory-efficient as well;
see details in \cite{huang2019globalsip} for an implementation for NMF.

\section{Stochastic Approaches}
Batch algorithms such as BCD and GN could have serious memory and computational issues, especially when the data tensor or matrix is large and dense.
Recall that the MTTKRP (i.e., $\H_n^\T\X_n$) costs ${\cal O}(R\prod_{n=1}^N I_n)$ operations, if no structure of the tensor can be exploited. This is quite expensive for large $I_n$ and high-order tensors. 
For big data problems, stochastic optimization is a classic workaround for avoiding memory/operation explosion. In a nutshell, stochastic algorithms are particularly suitable for handling problems having the following form:
\begin{equation}\label{eq:emprical}
    \min_{\bm \theta}~\frac{1}{L}\sum_{\ell=1}^L f_\ell(\bm \theta) + h(\bm \theta),
\end{equation}
where the first term is often called the ``empirical risk'' function in the literature.
The classic stochastic proximal gradient descent (SPGD) updates the optimization variables via
\begin{equation}\label{eq:stopgd}
 \bm \theta^{(t+1)}\leftarrow {\sf Prox}_h\left(\bm \theta^{(t)} - \alpha^{(t)} \bm g(\bm \theta^{(t)}) \right), 
\end{equation}    
where $\bm g(\bm \theta^{(t)}) $ is a random vector (or, ``stochastic oracle'') evaluated at $\bm \theta^{(t)}$, constructed through a random variable (RV) $\xi^{(t)}$.
The idea is to use an easily computable stochastic oracle to approximate the computationally expensive full gradient $\nabla f(\bm \theta)=\frac{1}{L}\sum_{\ell=1}^L \nabla f(\bm \theta)$, so that \eqref{eq:stopgd} serves as an economical version of the PGD algorithm.
A popular choice is $\bm g(\bm \theta^{(t)}) =\nabla f_\ell(\bm \theta^{(t)})$, where $\ell\in\{1,\ldots,L\}$ is randomly selected following the probability mass function (PMF) ${\sf Pr}(\xi^{(t)}=\ell)=1/L$. This simple construction has a nice property: $\bm g(\bm \theta^{(t)}) $ is an unbiased estimator for the full gradient given the history of random sampling, i.e., $$\nabla f(\bm \theta^{(t)})=\frac{1}{L}\sum_{\ell=1}^L \nabla f_\ell(\bm \theta^{(t)})=\mathbb{E}_{\xi^{(r)}}\left[  \bm g(\bm \theta^{(t)}) | {\cal H}^{(t)} \right],$$ 
where ${\cal H}^{(t)}$ collects all the RVs appearing before iteration~$t$.
The unbiasedness is often instrumental in establishing convergence of stochastic algorithms\footnote{Biased stochastic oracle and its convergence properties are also discussed in the literature; see, e.g., \cite{xu2015block}. However, the analysis is more involved. In addition, some conditions (e.g., bounded bias) are not easy to verify.}.
Another very important aspect is the variance of $\bm g(\bm \theta^{(t)}) $. Assume that the variance is bounded, i.e.,  $\mathbb{V}\left[  \bm g(\bm \theta^{(t)}) | {\cal H}^{(t)} \right] \leq \tau.$
Naturally, one hopes $\tau$ to be small---so that the average deviation of $\bm g(\bm \theta^{(t)})$ from the full gradient is small---and thus the SPGD algorithm will behave more like the PGD algorithm. 
Smaller $\tau$ can be obtained via using more samples to construct  $\bm g(\bm \theta^{(t)})$, e.g., using 
$$ 
\bm g(\bm \theta^{(t)})=\frac{1}{|{\cal B}^{(t)}|}\sum_{\ell\in {\cal B}^{(t)} }\nabla f_\ell(\bm \theta^{(t)}),
$$ 
where ${\cal B}^{(t)}$ denotes the index set of the $f_\ell(\bm \theta^{(t)})$'s sampled at iteration $t$. This leads to the so-called ``mini-batch'' scheme. Note that if $|{\cal B}^{(t)}|=L$, then $\tau=0$ and SPGD becomes the PGD. 
As we have mentioned, a smaller $\tau$ would make the convergence properties of SPGD more like the PGD, and thus is preferred.
However, a larger $|{\cal B}^{(t)}|$ leads to more operations for computing the stochastic oracle. 
In practice, this is a tradeoff that oftentimes requires some tuning to balance.

The randomness of stochastic algorithms makes characterizing the convergence properties of any single instance not meaningful.
Instead, the ``expected convergence properties'' are often used. For example, when $h(\bm \theta)$ is absent,  a convergence criterion of interest is expressed as follows:
\begin{equation}\label{eq:totalexpect}
     \liminf_{t\rightarrow \infty}{\mathbb{E}\left[ \left\|\nabla f\left(\bm \theta^{(t)}\right)\right\|_2^2 \right]}= 0, 
\end{equation}   
where the expectation is taken over all the random variables that were used for constructing the stochastic oracles for all the iterations (i.e., the ``total expectation").
Equation~\eqref{eq:totalexpect} means that every limit point of $\{ \bm \theta^{(t)} \}$ is a stationary point {\it in expectation}. When $h(\bm \theta)$ is present, similar ideas are utilized. Recall that $\bm 0\in \partial F(\bm \theta^{(t)})$ is the necessary condition for attaining a stationary point [cf.\ Eq.~\eqref{eq:convergence}].
In \cite{xu2015block}, the expected counterpart of \eqref{eq:convergence}, i.e., 
\begin{equation}\label{eq:constrainedexpct}
\liminf_{t\rightarrow \infty}\mathbb{E}[{\rm dist}(\bm 0, \partial F(\bm \theta^{(t)}) ) ]=0
\end{equation}  
is employed for establishing the notion of stationary-point convergence for nonconvex nonsmooth problems under the stochastic settings.
For both \eqref{eq:totalexpect} and \eqref{eq:constrainedexpct}, when some more assumptions hold (e.g., the solution sequence is bounded), the ``$\inf$'' notation can be removed, meaning that the whole sequence converges to a stationary point on average.

\subsection{Entry Sampling}
Many SLRD problems can be re-expressed in a similar form as that in \eqref{eq:emprical}.
One can rewrite the constrained CPD problem under the least squares fitting loss as follows:
\begin{equation}\label{eq:cpd_entry}
\begin{aligned}
\min_{\{\A_{n}\}_{n=1}^N}&~\frac{1}{L}\sum_{i_1=1}^{I_1}\ldots \sum_{i_N=1}^{I_N}{f_{i_1,\ldots,i_N}(\bm \theta)}+{h(\bm \theta)}
\end{aligned}
\end{equation}
where $L=\prod_{n=1}^N I_n$, $h(\bm \theta)={\sum_{n=1}^Nh_n(\A_n)}$ and $f_{i_1,\ldots,i_N}={( {\cal T}(i_1,\ldots,i_N)-\sum_{r=1}^R \prod_{n=1}^N \A_n(i_n,r))^2}$.
Assume ${\cal B}^{(t)}$ is a set of indices of the tensor entries that are randomly sampled [see Fig.~\ref{fig:sampling-strategies} (left)].
The corresponding SPGD update is as follows: $\bm \theta^{(t+1)}$ is given by 
\begin{align}\label{eq:spgd}
   {\sf Prox}_h\bigg(\bm \theta^{(t)} -\frac{\alpha^{(t)}}{|{\cal B}^{(t)}|} \sum_{(i_1,\ldots,i_N)\in{\cal B}^{(t)}} \nabla f_{i_1,\ldots,i_N}\big(\bm \theta^{(t)}\big) \bigg).  
\end{align}
It is not difficult to see that many entries of $\nabla f_{i_1,\ldots,i_N}(\bm \theta^{(t)})$ are zero, since $\nabla f_{i_1,\ldots,i_N}(\bm \theta^{(t)})$ only contains the information of $\A_{n}(i_n,:)$; we have
$\left[\nabla f_{i_1,\ldots,i_N}(\bm \theta^{(t)})\right]_g = 0$ for all $\theta_g \notin \{ \A_n(i_n,r)~|~(i_1,\ldots,i_N)\in{\cal B}^{(t)}\}$.  
The derivative w.r.t. $\A_n(i_n,:)$ for the sampled indices is easy to compute; see \cite{sidiropoulos2017tensor}.
This is essentially the idea in \cite{beutel2014flexifact} for coupled tensor and matrix decompositions. This kind of sampling strategy ensures that the constructed stochastic oracle is an unbiased estimation for the full gradient, and features very lightweight updates. Computing the term $\sum_{{\cal B}^{(t)}}\nabla f_{i_1,\ldots,i_N}(\bm \theta^{(t)})$ requires only ${\cal O}(R|{\cal B}^{(t)}|)$ operations, instead of ${\cal O}(R\prod_{n=1}^N I_n )$ operations for computing the full gradient.

\definecolor{basecolor}{RGB}{0,85,165}

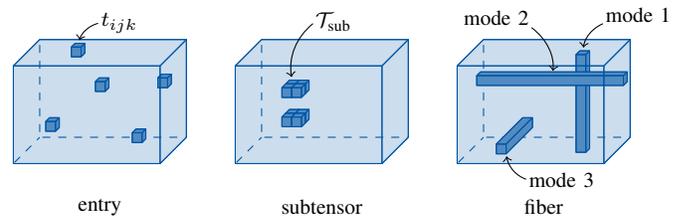
\begin{figure}[htp]
	\centering
	\begin{tikzpicture}[%
	all tensors/.append style={tensor scale=0.13, 2d, anchor=A},
	elm/.style={tensor, dim={1,1,1},fill=basecolor!60!white,draw=basecolor},
	pil/.style={ <-, shorten <=2pt,  shorten >=4pt,}]
	
	\node [tensor, dim={10,15,7}, only back faces, dashed back lines,draw=basecolor] (T1) {};
	\node at (T1.3 15 3) [elm] {};
	\node at (T1.1 5 6) [elm] (t11){};
	\node at (T1.3 9 2) [elm] {};
	\node at (T1.9 2 7) [elm] {};
	\node at (T1.9 12 4) [elm] {};
	\node at (T1.A) [tensor, dim={10,15,7}, anchor=A, fill opacity=0.2,
	fill=basecolor, draw=basecolor] (T1) {};
	
	\draw [pil,shorten >=1pt] ($(t11.A)!0.5!(t11.G)$) to [pil,bend left] ++(0.3,0.3) node
	[font=\footnotesize,anchor=west, inner sep=0pt] {$t_{ijk}$};
	
	\node at (T1.C) [tensor, dim={10,15,7}, only back faces, dashed back
	lines,draw=basecolor, anchor=A, xshift=1cm] (T2) {};
	\node at (T2.7 5 4) [elm] {};
	\node at (T2.7 6 4) [elm] {};
	\node at (T2.4 5 4) [elm] (t21) {};
	\node at (T2.4 6 4) [elm] {};
	\node at (T2.4 5 3) [elm] {};
	\node at (T2.4 6 3) [elm] {};
	\node at (T2.7 5 3) [elm] {};
	\node at (T2.7 6 3) [elm] {};
	\node at (T2.A) [tensor, dim={10,15,7}, anchor=A, fill opacity=0.2,
	fill=basecolor, draw=basecolor] (T2) {};
	
	\draw [pil,shorten >=1pt] ($(t21.A)!0.5!(t21.G)$) to [pil,bend left] ++(0.3,0.8) node
	[font=\footnotesize,anchor=west, inner sep=0pt] {$\ten{T}_{\text{sub}}$};

	\node at (T2.C) [tensor, dim={10,15,7}, only back faces, dashed back
	lines,draw=basecolor, anchor=A, xshift=1cm] (T3) {};
	\node at (T3.1 12 4) [elm,dim={10,1,1}] (t31) {};
	\node at (T3.2 3 1) [elm,dim={1,15,1}] (t32) {};
	\node at (T3.9 5 1) [elm,dim={1,1,7}] (t33) {};
	\node at (T3.A) [tensor, dim={10,15,7}, anchor=A, fill opacity=0.2,
	fill=basecolor, draw=basecolor] (T3) {};
	
	\draw [pil,shorten >=1pt] ($(t31.A)!0.5!(t31.G)$) to [pil,bend left] ++(0.3,0.4) node
	[font=\footnotesize,anchor=south west, inner sep=0pt] {mode 1};
	
	\draw [pil,shorten <=3pt,shorten >=1pt] (t32) to [pil,bend right] ++(-0.3,0.7) node
	[font=\footnotesize,anchor=south east,inner sep=0pt] {mode 2};
	
	\draw [pil,shorten <=1pt, shorten >=1pt] (t33.D) to [pil,bend right] ++(0.3,-0.35) node
	[font=\footnotesize,anchor=west,inner sep=0pt] {mode 3};
	
	\node at (T1) [yshift=-1.4cm,font=\footnotesize] {entry};
	\node at (T2) [yshift=-1.4cm,font=\footnotesize] {subtensor};
	\node at (T3) [yshift=-1.4cm,font=\footnotesize] {fiber};
	
	\end{tikzpicture}
	\vspace{-5mm}
	\caption{Various sampling strategies used in stochastic optimization
		algorithms.}
	\label{fig:sampling-strategies}
	\vspace{-.5cm}
\end{figure}


\subsection{Subtensor Sampling}
Entry-sampling based approaches are direct applications of the conventional PSGD for tensor decomposition. However, these methods do not leverage existing tensor decomposition tools. 
One way to take advantage of existing tensor decomposition algorithms is sampling subtensors, instead of entries.
The {\em randomized block sampling} (RBS) algorithm \cite{vervliet2016randomized} considers the unconstrained CPD problem. The algorithm samples a subtensor 
$$
{\cal T}_{\rm sub}^{(t)}={\cal T}({\cal S}_1,\ldots,{\cal S}_N) \approx \sum_{r=1}^R \A_1({\cal S}_1,r)\circ \ldots \circ \A_N({\cal S}_N,r) 
$$ 
at every iteration $t$ and updates the latent variables by computing \emph{one} optimization step using:
\begin{equation}\label{eq:RBS}
\begin{aligned}
\bm \theta_{\rm sub}^{(t+1)} &\leftarrow \arg\min_{\bm \theta_{\rm sub}} \left\|{\cal T}_{\rm sub}^{(t)}-\llbracket \A_1^{\rm sub}, \ldots, \A_N^{\rm sub}\rrbracket\right\|^2_{\rm F},\\
\bm \theta_{\rm -sub}^{(t+1)}& \leftarrow \bm \theta_{\rm -sub}^{(t)},
\end{aligned}
\end{equation}
where all variables affected by ${\cal T}_{\rm sub}^{(t)}$ are collected in $\bm \theta_{\rm sub}=[{\rm vec}(\A_1^{\rm sub});\ldots,{\rm vec}(\A_N^{\rm sub})]$, $\A_n^{\rm sub}=\A_n({\cal S}_n,:)$, and $\bm \theta_{\rm -sub}$ contains all the other optimization variables. As each update in \eqref{eq:RBS} involves one step in a common tensor decomposition problem, many off-the-shelf algorithms, such as ALS or GN, can be leveraged \cite{vervliet2016randomized}.

The above algorithm works well, especially when the tensor rank is low and the sampled subtensors already have identifiable latent factors---under such cases, the estimated $\A_n^{\rm sub}$ from subtensors can serve as a good estimate for the corresponding part of $\A_n$ after one or two updates. 
In practice, one needs not to exactly solve the subproblems in \eqref{eq:RBS}. Combining with some trust region considerations, the work in \cite{vervliet2016randomized} suggested using a one-step GN or one-step regularized ALS to update $\bm \theta_{\rm sub}$. 
Note the sampled subtensors
are typically not independent under this framework, since one wishes to update every unknown parameter in an equally frequent way; see~\cite{vervliet2016randomized}. This is quite different from established conventions in stochastic optimization, which makes convergence analysis for RBS more challenging than the entry sampling based methods.

\subsection{Fiber Sampling}
In principle, the entry sampling and SPGD idea in \eqref{eq:spgd} can handle any $h(\cdot)$ that admits simple proximal operators.
In addition, the RBS algorithm can be applied together any constraint compatible with the GN framework as well.
However, such sampling strategies are no longer viable when it comes to constraints/regularizers that are imposed on the columns of the latent factors, e.g., 
the probability simplex constraint that is often used in statistical learning $(\bm 1^\T\A_n =\bm 1^\T,~\bm A_n\geq \bm 0)$, the constraint
$\|\A_n\|_{2,1}=\sum_{i_n=1}^{I_N}\|\A_n(i_n,:)\|_2$ used for promoting row-sparsity, or the total variation/smoothness regularization terms on the columns of $\A_n$. 
The reason is that ${\cal T}_{\rm sub}$ only contains information of $\A_n({\cal S}_n,:)$---which means that enforcing column constraints on $\bm A_n$ is not possible if updates in \eqref{eq:spgd} or \eqref{eq:RBS} are employed. 

Recently, the works in \cite{battaglino2018practical,fu2019block} advocate to sample a (set of) mode-$n$ ``fibers'' for updating $\A_n$. A mode-$n$ fiber of the tensor ${\cal T}$ is an $I_n$-dimensional vector that is obtained by varying the mode-$n$ index while fixing others of ${\cal T}$ [see Fig.~\ref{fig:sampling-strategies} (right)]. The interesting connection here is that
\begin{align*}
\underbrace{{\cal T}(i_1,\ldots,i_{n-1},:,i_{n+1},\ldots,i_N)}_{\text{a mode-$n$ fiber}}
=\X_n(j_n,:),
\end{align*}
where $j_n=1+\sum_{\ell=1,\ell\neq n}(i_\ell-1)J_\ell$ and $J_\ell=\prod_{m=1,m\neq n}^{\ell-1}I_m$. Under this sampling strategy, the whole $\A_n$ can be updated in one iteration. Specifically, in iteration $t$, the work in \cite{battaglino2018practical} updates $\A_n$ for $n=1,\ldots,N$ sequentially, as in the BCD case. To update $\A_n$, it samples a set of mode-$n$ fibers, indexed by ${\cal Q}_n^{(t)}$ and solve a `sketched least squares' problem:
\begin{equation}
\min_{\A_n}~\left\| \X_n({\cal Q}_n^{(t)},:) - \H_n^{(t)}({\cal Q}_n^{(t)},:)\A_n^\T \right\|_{\rm F}^2,
\end{equation}
whose solution is $$ \A_n^{(t+1)} \leftarrow ( \H_n^{(t)}({\cal Q}_n^{(t)},:)^{\dagger}\X_n({\cal Q}_n^{(t)},:))^\T. $$
This simple sampling strategy makes sure that every entry of $\A_n$ can be updated in iteration $t$.
The rationale behind is also reasonable: If the tensor is low-rank, then one does not need to use all the data to solve the least squares subproblems---using randomly sketched data is enough, if the system of linear equations $\X_n = \H^{(t)}({\cal Q}_n^{(t)},:)\A_n^\T$ is over-determined, it returns the same solution as solving $\X_n = \H^{(t)}\A_n^\T$. 

The work in \cite{battaglino2018practical} did not explicitly consider structural information on $\A_n$'s, and the convergence properties of the approach are unclear.
To incorporate structural information and to establish convergence, the recent work in \cite{fu2019block} offered a remedy. There, a block-randomized sampling strategy was proposed to help establish unbiasedness of the gradient estimation. Then, PGD is combined with fiber sampling for handling structural constraints.
The procedure consists of two sampling stages: first, randomly sample a mode $n\in\{1,\ldots,N\}$ with random seed $\zeta^{(t)}$ such that ${\sf Pr}(\zeta^{(t)} = n)=1/N$. Then, sample a set of mode-$n$ fibers indexed by ${\cal Q}_n$ uniformly at random (with another random seed $\xi^{(t)}$).  Using the sampled data, construct
\begin{equation}\label{eq:Gn}
 \G^{(t)} =[\G^{(t)}_1;\ldots;\G^{(t)}_N],
\end{equation}     
where $\G_n^{(t)} =  \A_n\B^\T\B-\X_n({\cal Q}_n,:)^\T\B$ with 
$\B = \H^{(t)}({\cal Q}_n^{(t)},:)$, and $\G_k^{(t)}=\bm 0$ for $k\neq n$.  
This block-randomization technique entails the following equality:
\begin{equation}\label{eq:brascpd_g}
{\rm vec}\left(\mathbb{E}_{\xi^{(t)}}\left[ \G^{(t)}|{\cal H}^{(t)},\zeta^{(t)} \right]\right)=c\nabla f(\bm \theta^{(t)}),
\end{equation}
where $c>0$ is a constant; i.e., the constructed stochastic vector is an unbiased estimation (up to a constant scaling factor) for the full gradient, conditioned on the filtration. Then, the algorithm updates the latent factors via
\begin{equation}\label{eq:brascpd}
\A_n^{(t+1)} \leftarrow {\sf Prox}_{h_n}\left( \A_n^{(t)} -\alpha^{(t)} \G_n^{(t)} \right).
\end{equation}
Because of \eqref{eq:brascpd_g}, the above is almost identical to single block SPGD, and thus enjoys similar convergence guarantees \cite{fu2019block}.

Fiber sampling approaches as in \cite{battaglino2018practical} and \cite{fu2019block} are economical, since they never need to instantiate the large matrix $\H_n$ or to compute the full MTTKRP. 
{ A remark is that fiber sampling is also of interest in partially observed tensor recovery \cite{kanatsoulis2019regular,zhang2020spectrum}; in Section \ref{sec:tractable} it will actually be argued that under mild conditions exact completion of a fiber-sampled tensor is possible via a matrix eigenvalue decomposition \cite{sorensen2015fibersampling}.}


\subsection{Adaptive Step-size Scheduling}
Implementing stochastic algorithms oftentimes requires somewhat intensive hands-on tuning for selecting hyperparameters, in particular, the step size $\alpha^{(t)}$. Generic SGD and SPGD analyses suggest to set the step size sequence following the Robbins and Monro's rule, i.e.,
$\sum_{t=0}^\infty \alpha^{(t)} =\infty,~\sum_{t=0}^\infty (\alpha^{(t)})^2<\infty.$ 
The common practice is to set $\alpha^{(t)} =\alpha/t^\beta$ with $\beta>1$, but the ``best'' $\alpha$ and $\beta$ for different problem instances can be quite different. 
To resolve this issue, adaptive step-size strategies that can automatically determine $\alpha^{(r)}$ are considered in the literature. The RBS method in \cite{vervliet2016randomized} and the fiber sampling method in \cite{fu2019block} both consider adaptive step-size selection for tensor decomposition.
In particular, the latter combines the insight of \texttt{adagrad} that has been popular in deep neural network training together with block-randomized tensor decomposition to come up with an adaptive step-size scheme (see ``{\bf Adagrad for Stochastic SLRD}").

\begin{figure}
	\centering
	\setlength{\figurewidth}{6cm}
    \setlength{\figureheight}{3cm}
    \input{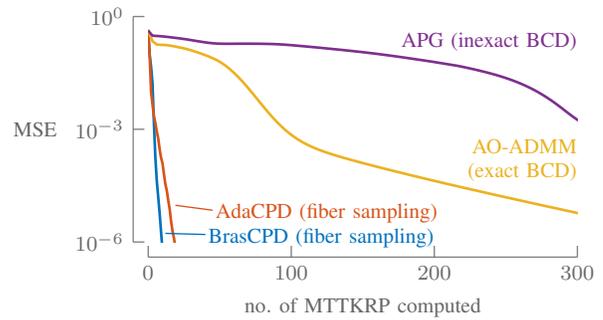}
	\caption{Stochastic algorithms [BrasCPD (manually fine-tuned step size) and AdaCPD (adaptive step size)] use significantly fewer operations to reach a good estimation accuracy for the latent factors, compared to batch algorithms. The MSE for estimating the $\A_{n}$'s against the number of full MTTKRP used. The CP rank is 10 and $I_n=100$ for all $n=1,2,3$. Figure reproduced from \cite{fu2019block}. Permission will be sought upon publication.} 
	\label{fig:fig1}
	\vspace{-,5cm}
\end{figure}

\begin{mdframed}[backgroundcolor=blue!10,topline=false,
	rightline=false,
	leftline=false,
	bottomline=false]

\noindent
{\bf Adagrad for Stochastic SLRD}
In \cite{fu2019block}, the following term is updated for each block $n$ under the block-randomized fiber sampling framework:
\begin{align*}
[\bm \eta^{(t)}_{n}]_{i,r} &\leftarrow \frac{1}{\left(b + \sum_{q=1}^{t}[{\bm G}_{(n)}^{(q)}]_{i,r}^2\right)^{1/2+\epsilon}},\label{eq:etaada}
\end{align*}
where $b$ and $\epsilon$ are inconsequential small positive quantities for regularization purpose.
Then, the selected block is updated via
\begin{equation}\label{eq:adas=cpd}
\A_n^{(t+1)} \leftarrow {\sf Prox}_{h_n}\left( \A_n^{(t)} -\bm \eta^{(t)}_n \circledast \G_n^{(t)} \right).
\end{equation}
The above can be understood as a data-adaptive pre-conditioning for the stochastic oracle $\G_n^{(t)}$. Implementing \texttt{adagrad} based stochastic CPD (AdaCPD) is fairly easy, but in practice it often saves a lot of effort for fine-tuning $\alpha^{(t)}$ while attaining competitive convergence speed; see Fig.~\ref{fig:fig1}.
This also shows the potential of adapting the well-developed stochastic optimization tools in deep neural network training to serve the purpose of SLRD.

It is shown in \cite{fu2019block} that, using the \texttt{adagrad} version of the fiber sampling algorithm, every limit point of $\{\bm \theta^{(t)}\}$ is a stationary point in expectation, if $h(\bm \theta)$ is absent. However, convergence in the presence of nonsmooth $h(\bm \theta)$ is still an open challenge.  
\end{mdframed}



%
%
%
%
%
%
%
%
%
%
%
%
%
%
%
%
%
%
%
%
%

\smallskip

In Fig.~\ref{fig:fig1}, we show the MSE on the estimated $\A_{n}$'s obtained by different algorithms after using a certain number of full MTTKRP (which serves as a unified complexity measure). Here, the tensor has size $100\times 100 \times 100$ and its CP rank is $R=10$. One can see that stochastic algorithms (BrasCPD and AdaCPD) work remarkably well in this simulation. In particular, the adaptive step size algorithm exhibits promising performance without tuning step-size parameters. We also would like to mention that the stochastic algorithms naturally work with incomplete data (e.g., data with missing entries or fibers), since the updates only rely on partial data.

\bigskip

{ 
Table~\ref{tab:handbook} presents an incomplete summary of structural constraints/regularization terms (together with the Euclidean data fitting-based CPD cost function) that can be handled by the introduced nonconvex optimization frameworks. One can see that different frameworks may be specialized for different types of structural constraints and regularization terms. 
In terms of accommodating structural requirements, the AO-ADMM algorithm \cite{huang2016flexible} and the GN framework offered in Tensorlab \cite{vervliet2016tensorlab3} may be the most flexible ones, since they can handle multiple structural constraints simultaneously. 
}

\begin{table*}[htbp]
 \centering
   \begin{threeparttable}[b]

  \caption{An incomplete summary of structural constraints that can be handled by some representative algorithms.}  \label{tab:handbook}%
				\centering
    \begin{tabular}{lcccc}
    \toprule
    Structural constraint or regularization & $\begin{subarray}{c} \text{AO-ADMM~\cite{huang2016flexible}} \\ \text{(exact BCD)}\end{subarray}$  & $\begin{subarray}{c} \text{APG~\cite{xu2013block}} \\ \text{(inexac BCD)}\end{subarray}$ &  $\begin{subarray}{c} \text{Tensorlab~\cite{vervliet2016tensorlab3}} \\ \text{(GN)}\end{subarray}$  & $\begin{subarray}{c} \text{AdaCPD~\cite{fu2019block}} \\ \text{(stochastic)}\end{subarray}$ \\
    \midrule
    Nonnegativity ($\A_n\geq \bm 0$) & \CheckmarkBold     & \CheckmarkBold     & \CheckmarkBold      & \CheckmarkBold  \\
    Sparsity ($\|\A_n\|_1=\sum_{i=1}^{I_n}\sum_{r=1}^R |\A_n(i,r)|$)    & \CheckmarkBold     & \CheckmarkBold      & \CheckmarkBold\tnote{+}      & \CheckmarkBold  \\
    Column group  sparsity ($\|\A_n\|_{2,1}=\sum_{r=1}^R\|\A_n(:,r)\|_2$)    & \CheckmarkBold     & \CheckmarkBold      & \CheckmarkBold\tnote{+}      & \CheckmarkBold  \\
    Row group  sparsity ($\|\A_n^\T\|_{2,1}=\sum_{i=1}^{I_n}\|\A_n(i,:)\|_2$)    & \CheckmarkBold     & \CheckmarkBold      & \CheckmarkBold\tnote{+}      & \CheckmarkBold  \\
    Total variation ($\|\bm T_1\bm A_n\|_1$)\tnote{*} & \CheckmarkBold      & \CheckmarkBold      & \CheckmarkBold\tnote{+}      & \CheckmarkBold  \\
    Row prob. simplex ($\A_n\bm 1=\bm 1,~\A_n\geq \bm 0$) & \CheckmarkBold      & \CheckmarkBold      & \CheckmarkBold      & \CheckmarkBold  \\
    Column prob. simplex ($\bm 1^\T\A_n=\bm 1^\T,~\A_n\geq \bm 0$) & \CheckmarkBold      & \CheckmarkBold      & \CheckmarkBold      & \CheckmarkBold \\
    Tikhonov smoothness ($\|\bm T_2\A_n\|_{\rm F}^2$)\tnote{*} & \CheckmarkBold     & \CheckmarkBold      & \CheckmarkBold      & \CheckmarkBold  \\
    Decomposition symmetry ($\A_n=\A_m$)&       &       & \CheckmarkBold      &  \\
    Boundedness ($a\leq \A_n(i,r)\leq b$) & \CheckmarkBold      & \CheckmarkBold     & \CheckmarkBold      & \CheckmarkBold  \\
    Coupled factorization (see \cite{ibrahim2019crowdsourcing,ibrahim2019stochastic,sorensen2015coupledcanonicalpart1,kanatsoulis2019regular}) & \CheckmarkBold      & \CheckmarkBold     & \CheckmarkBold      & \CheckmarkBold  \\
    Multiple structures combined (e.g., $\|\bm T_2\A_n\|_1 + \|\A_n\|_{2,1}$) & \CheckmarkBold      &       &  \CheckmarkBold       &  \\
    \bottomrule
    \end{tabular}
    
      \begin{tablenotes}
     \item[*] The operators $\bm T_1$ and $\bm T_2$ are sparse circulant matrices whose expressions can be found in the literature, e.g., \cite{CVX2004}.
     \item[+] GN-based methods (except for ProxGN in \cite{huang2019globalsip}) work with differentiable functions. In Tensorlab, the $\ell_1$ norm-related non-differentiable terms are handled using function-smoothing techniques as approximations; see details in \cite{vervliet2018numericaloptimization}.
   \end{tablenotes}
  \end{threeparttable}

\end{table*}%

\section{More Discussions and Conclusion} \label{sec:morediscuss}


\subsection{Exploiting Structure at Data Level}\label{sec:orgd2a5b2a}

Until now, the focus has been on exploiting the multilinear structure of the
decomposition to come up with scalable SLRD algorithms. 
In many cases the tensor itself has additional structure that can be exploited to reduce complexity of some ``bottleneck operations'' such as MTTKRP (which is used in both GN and BCD) or computing the fitting residual (needed in GN). 
Note that for batch algorithms, both computational and memory complexities of these operations scale as $\order{\text{tensor entries}}$. 
%
For classic methods like BCD, there is rich literature on exploiting data structure, in particular sparsity, to avoid memory or flop explosion; see \cite{kolda2009tensor,sidiropoulos2017tensor} and references therein.
For all batch methods, it is crucial to exploit data structure in
order to reduce the complexity of computing \(\bm f\) and  \(\bm{g}\) to 
$\order{\text{parameters in representation}}$. The key
is avoiding the explicit construction of the residual $\ten{F}$. The techniques
for second-order methods and constraints outlined in Sec.~\ref{sec:orgd2a5b2a} can be used without changes, as the computation of the Gramian as well as the Jacobians
\(\tilde{\bm{J}}\) resulting from parametric, symmetry or coupling constraints
are independent of the tensor \cite{vervliet2016exploitingefficient}, which can
be verified from~\eqref{eq:system-gn}. This way the nonnegative CPD of GB
size tensors, or deterministic BSS problems with up to millions of samples can
be handled easily on simple laptops or desktops; see
\cite{vervliet2016exploitingefficient} for examples.

\subsection{Other Loss Functions} 

In the previous sections, we have focused on the standard Euclidean distance to measure the error of the data fitting term. This is by no means the best choice in all scenarios. It corresponds to the { MLE} assuming the input tensor is a low-rank tensor to which additive i.i.d.\ Gaussian noise is added. 
It may be crucial  in some cases to adopt other data fitting terms. Let us mention { an array of} important examples: 

\noindent $\bullet$ For count data, such as documents represented as vectors of word counts (this is the so-called bag-of-words model), 
the matrix/tensor is nonnegative and typically sparse (most documents do not use most words from the dictionary) for which Gaussian noise is clearly not appropriate.  
Let us focus on the matrix case for simplicity. If we assume the noise added to the entry $(i,j)$ of the input matrix $\X$ is Poissonian of parameter $\lambda = (\A_1\A_2)_{i,j}$, we have  
$
{\sf Pr}( X_{i,j} = k ) =  \nicefrac{e^{-\lambda}  \lambda^k}{k!}
$
 with $k \in \mathbb{Z}_+$.  
The MLE leads to minimizing the KL divergence between $\X$ and $\A_1\A_2$: 
\begin{equation} \label{eq:KLmle}
\min_{\A_1, \A_2} \sum_{i,j} \X_{i,j} \log \frac{\X_{i,j}}{(\A_1\A_2^\T)_{i,j}} - \X_{i,j} + (\A_1\A_2^\T)_{i,j}.
\end{equation} 
{ The KL divergence is also widely used in imaging because the acquisition can be seen as a photon-counting process (note that, in this case, the input matrix is not necessarily sparse).
} 

{ 
\noindent $\bullet$ Multiplicative noise, for which each entry of the low-rank tensor is multiplied with some noise, has been shown to be particularly well adapted to audio signals. For example, if the multiplicative noise follows a Gamma distribution, the MLE minimizes the Itakura-Saito (IS) divergence between the observed tensor and its low-rank approximation~\cite{fevotte2009nonnegative}; in the matrix case with $\X \approx \A_1\A_2^\T$, it is given by 
\begin{equation} \label{eq:ISmle}
\min_{\A_1, \A_2} 
\sum_{i,j} 
\frac{\X_{i,j}}{(\A_1\A_2^\T)_{i,j}}  
- \log \frac{\X_{i,j}}{(\A_1\A_2^\T)_{i,j}} 
- 1 . 
\end{equation} 
}

\noindent $\bullet$  In the presence of outliers, that is, the noise has some entries with large magnitude, 
using the component-wise $\ell_1$-norm is more appropriate 
\begin{equation} \label{eq:L1mle}
\sum_{i_1, i_2,\dots,i_N}  \left| {\cal T}(i_1,\ldots,i_N)-\sum_{r=1}^R \prod_{n=1}^N \A_n(i_n,r) \right|, 
\end{equation} 
and corresponds to the MLE for Laplace noise { ~\cite{vorobyov2005robust}}. 
This is closely related to robust PCA and can be used for example to extract the low-rank background from moving objects (treated as outliers) in a video sequence~\cite{gillis2018complexity}.  
{ When ``gross outliers'' heavily corrupt a number of slabs of the tensor data (or columns/rows of the matrix data), optimization objectives involving nonconvex mixed $\ell_2/\ell_p$ functions (where $0<p\leq 1$) may also be used~\cite{fu2015joint,fu2016robust}. For example, the following fitting cost may be used when one believes that some columns of $\X$ are outliers \cite{fu2016robust}:
\[  \sum_{i_2=1}^{I_2} \left\| \X(:,i_2) - \A_1\A_2(i_2,:)^\T \right\|_2^p,      \]
where $0<p\leq 1$ is used to downweight the impact of the outlying columns. 
}

{ 
\noindent $\bullet$ For quantized signals, that is, signals whose entries have been rounded to some accuracy, an appropriate noise model is the uniform distribution\footnote{ For the $\ell_{\infty}$ norm to correspond to the MLE, all entries must be rounded with the same absolute accuracy (e.g., the nearest integer), which is typically not the case in most programming languages.}. 
For example, if each entry of a low-rank matrix are rounded to the nearest integer, then each entry of the noise can be modeled with the uniform distribution in the interval $[-0.5,0.5]$. 
The corresponding MLE minimizes the component-wise $\ell_\infty$ norm; replacing $\sum_{i_1, i_2,\dots,i_N}$ by  
$\max_{i_1, i_2,\dots,i_N}$ in~\eqref{eq:L1mle}.  
} 

\noindent $\bullet$ If the noise is not identically distributed among the entries of the tensor, a weight should be assigned to each entry. For example, for independently distributed Gaussian noise, the MLE minimizes  
\[
\sum_{i_1=1}^{I_1}\ldots \sum_{i_N=1}^{I_N} 
\frac{\left( {\cal T}(i_1,\ldots,i_N)-\sum_{r=1}^R \prod_{n=1}^N \A_n(i_n,r) \right)^2} 
{\sigma^2(i_1,\dots,i_N)}, 
\] 
 where $\sigma^2(i_1,\dots,i_N)$ is the variance of the noise for the entry at position $(i_1,\dots,i_N)$. 
 Interestingly, for missing entries, $\sigma(i_1,\dots,i_N) = +\infty$ corresponds to a weight of zero while, if there is no noise, that is, $\sigma(i_1,\dots,i_N) = 0$, the weight is infinite so that the entry must be exactly reconstructed.


\smallskip

In all cases above, we end up with more complicated optimization problems because the nice properties of the Euclidean distance are lost; in particular Lipschitz continuity of the gradient ({ the $\ell_1$ and $\ell_\infty$ norms} are even nonsmooth). 
For the weighted norm, the problem might become ill-posed (the optimal solution might not exist, even with nonnegativy constraints) in the presence of missing entries because some weights are zero so that the weighted ``norm'' is actually not a norm. 
{ For the KL and IS divergences, the gradient of the objective is not Lipschitz continuous, and the objective not defined everywhere: $\X_{i,j} > 0$ requires $(\A_1\A_2^\T)_{i,j} > 0$ in~\eqref{eq:KLmle} and~\eqref{eq:ISmle}.}  
The most popular optimization method { for these divergences} is multiplicative updates which is an inexact BCD { approach}; see Section~\ref{sec:inexactBCD}. 
For the componentwise $\ell_1$, { $\ell_\infty$ norms and nonconvex $\ell_2/\ell_p$ functions}, subgradient descent (which is similar to PGD), iteratively reweighed least squares, or exact BCD are popular approaches; { see, e.g., \cite{vorobyov2005robust,fu2016robust,fu2015joint}. Some of these objectives (e.g., the KL divergence and the component-wise $\ell_1$ norm) can also be handled under a variant of the AO-ADMM framework with simple updates but possibly high memory complexities \cite{huang2016flexible}.}
In all cases, convergence will be typically slower than for the Euclidean distance.

\subsection{Tractable SLRD Problems and Algorithms}\label{sec:tractable}
We have introduced a series of nonconvex optimization tools for SLRD that are all supported by stationary-point convergence guarantees. 
However, it is in general unknown if these algorithms will reach a globally optimal solution (or, if the LRDMs can be exactly found).
While convergence to the global optimum can be observed in practical applications, establishing pertinent theoretical guarantees is challenging given the NP-hardness of the problem, 
\cite{hillar2013most,vavasis2009complexity,anandkumar2014tensor1,gillis2018complexity}. Nevertheless, 
in certain settings the computation of LRDMs is known to be tractable. We mention the following:

\noindent
$\bullet$
In the case where a fully symmetric tensor admits a CPD with all latent factors identical and orthogonal (i.e., all the $\bm A_n$'s are identical and $\bm A_n^\T\A_n=\bm I$), the latent factors can be computed using
a power iteration/deflation-type algorithm \cite{anandkumar2014tensor1}. This is analogous to the computation of the eigendecomposition of a symmetric matrix through successive power iteration and deflation. A difference is that a symmetric matrix can be exactly diagonalized by an orthogonal eigentransformation, while a generic higher-order tensor can only approximately be  diagonalized; the degree of diagonalizability affects the convergence \cite{espig2015convergence}. 
By itself, CPD with identical and orthogonal $\A_n$'s is a special model that is not readily encountered in many applications.
However, in an array of blind source separation and machine learning problems (e.g., independent component analysis, topic modeling and community detection), it is under some conditions possible to transform higher-order statistics so that they satisfy this special model up to estimation errors. 
In particular, the second-order statistics can be used for a prewhitening that is guaranteed to orthogonalize the latent factors when the decomposition is exact. For deflation-based techniques that do not require orthogonality nor symmetry, see \cite{phan2015tensor,ever20eigenspace}.

\noindent
$\bullet$ 
Beyond CPD with identical and orthogonal latent factors, eigendecomposition-based algorithms have a long history for finding the exact CPD under various conditions. 
The simplest scenario is where two factor matrices have full column rank and the third factor matrix does not have proportional columns. In this scenario, the exact CPD can be found from the generalized eigenvalue decomposition of a pencil formed by two tensor slices (or linear combinations of slices) \cite{leurgans1993decomposition}. The fact that in the first steps of the algorithm the tensor is reduced to just a pair of its slices, implies some bounds on the accuracy, especially in cases where the rank is high compared to the tensor dimensions, i.e. when a lot of information is extracted from the two slices \cite{beltran2019pencil}. To mitigate this, \cite{ever20eigenspace} presents an algebraic approach in which multiple pencils are each partially used, in a way that takes into account their numerical properties. \\
Moreover, the working conditions of the basic eigendecomposition approach have been relaxed to situations in which 
only one factor is required to be full column rank \cite{delathauwer2006linkbetween}. The method utilizes a bilinear mapping to convert the more general CPD problem to the ``simplest scenario'' above. 
This line of work has been further extended to handle cases where the latent factors are all allowed to be rank deficient, enabling exact algebraic computation up to the famous Kruskal bound and beyond \cite{domanov2014canonicalpolyadic,domanov2017canonical}. Algorithms of this type have been proposed for other tensor decomposition models as well, e.g., block-term decomposition and LL1 decomposition \cite{de2008decompositions_3,domanov2020uniqueness}, coupled CPD \cite{sorensen2015coupledcanonicalpart2}, and CPD of incomplete fiber-sampled tensors \cite{sorensen2015fibersampling}. 
While the accuracy of these methods is sometimes limited in practical noisy settings, the computed results often provide good initialization points for the introduced iterative nonconvex optimization-based methods.  

\noindent
$\bullet$ 
In \cite{ever20existence} noise bounds are derived under which the CPD minimization problem is well-posed and the cost function has only one local minimum, which is hence global.

\noindent
$\bullet$ 
Many unconstrained low-rank matrix estimation problems (e.g., compressed matrix recovery and matrix completion) are known to be solvable via nonconvex optimization methods, under certain conditions \cite{chi2019nonconvex}.
Structure-constrained matrix decomposition problems are in general more challenging, but solvable cases also exist under some model assumptions.
For example, {\it separable NMF} tackles the NMF problem under the assumption that a latent factor contains a column-scaled version of the identity matrix as its submatrix. This assumption facilitates a number of algorithms that provably output the target latent factors, even in the noisy cases; see tutorials in \cite{gillis2014and,fu2018nonnegative}.
Solvable cases also exist in dictionary learning that identifies a sparse factor in an ``overcomplete'' basis. If the sparse latent factor is generated following a Gaussian-Bernoulli model, then it was shown that the optimization landscape under an ``inverse filtering'' formulation is ``benign''---i.e., all local minima are also global minima. Consequently, a globally optimal solution can be attained via nonconvex optimization methods \cite{sun2017complete}.

\subsection{Other Models}
The algorithm design principles can be generalized to cover other models, e.g., BTD, LL1, Tucker, and Tensor Train (TT)/hierarchical Tucker (hT), to name a few \cite{oseledets2008tucker,savas2010quasi,ishteva2011best,hackbusch2012tensor,grasedyck2013literature,khoromskij2018tensor}. 
Note that, in their basic form, BTD, LL1, Tucker and TT/hT involve subspaces rather than vectors, so that optimization on manifolds is a natural framework. 
Some extensions of SLRD are straightforward. For instance, both BCD and second-order algorithms for structured Tucker, BTD, and LL1 decompositions exist \cite{xu2013block,de2008decompositions_3,sorber2013optimizationbasedalgorithms,sorber2015structureddatafusion}. GN-based methods were also considered for nonnegativity-constrained Tucker decomposition. LL1 can be regarded as CPD with repeated columns in some latent factor matrices, and thus the parametrization techniques can be used to come up with GN algorithms for LL1, as constrained CPD \cite{vervliet2018numericaloptimization,sorber2013optimizationbasedalgorithms}.
However, some extensions may require more effort. For example, in stochastic algorithm design, different tensor models and structural constraints may require custom design of sampling strategies, as we have seen in the CPD case. This also entails many research opportunities ahead.

%

%


\subsection{Concluding Remarks}
In this article, we introduced three types of nonconvex optimization tools that are effective 
for SLRD. Several remarks are in order:

\noindent
$\bullet$ The BCD-based approaches are easy to understand and implement. The inexact BCD and extrapolation techniques are particularly useful in practice. This line of work can potentially handle a large variety of constraints and regularization terms, if the subproblem solver is properly chosen. The downside is that BCD is a first-order optimization approach at a high level. Hence, the speed of convergence is usually not fast. Designing effective and lightweight acceleration strategies may help advance BCD-based SLRD algorithms.

\noindent
$\bullet$ The GN-based approaches are powerful in terms of convergence speed and per-iteration computational complexity. They are also the foundation of the tensor computation infrastructure \texttt{Tensorlab}. On the other hand, the GN approaches
are specialized for NLS and smoothed objective functions. In other words,
they may not be as flexible as BCD-based approaches in terms of incorporating structural information. Using ProxGN may improve the flexibility, but the subproblems arising in the ProxGN framework are not necessarily easy to solve. Extending the second-order approaches to accommodate more structural requirements and objective functions other than the least squares loss promises a fertile research ground.

\noindent
$\bullet$ The stochastic approaches strike a balance between per-iteration computational/memory complexity and the overall decomposition algorithm effectiveness. Different sampling strategies may be able to handle different types of structural information. Stochastic optimization may involve more hyperparameters to tune (in particular, the mini-batch size and step size), and thus may require more attentive software engineering for implementation. Convergence properties of stochastic tensor/matrix decomposition algorithms are not as clear, which also poses many exciting research questions for the tensor/matrix and optimization communities to explore. 

\makeatletter
\let\c=\@c@old
\makeatother

\bibliographystyle{IEEEtran}

\bibliography{myabbrv,IEEEabrv,refs,refs_xiao}



\end{document}